\documentclass[pra,onecolumn,preprint]{revtex4}
\usepackage{graphicx}
\usepackage{amsfonts}

\begin{document}

\preprint{}
\title[Beyond Quantum Computation]{Optical analogy to quantum computation
based on classical fields modulated pseudorandom phase sequences}
\author{Jian Fu, Xutai Ma, Wenjiang Li and Shuo Sun}
\affiliation{State Key Lab of Modern Optical Instrumentation, College of Optical Science
and Engineering, Zhejiang University, Hangzhou, 310027, China}
\pacs{03.67.-a, 42.50.-p}

\begin{abstract}
We demonstrate that a tensor product structure and optical analogy of
quantum entanglement can be obtained by introducing pseudorandom phase
sequences into classical fields with two orthogonal modes. Using the
classical analogy, we discuss efficient simulation of several typical
quantum states, including product state, Bell states, GHZ state, and W
state. By performing quadrature demodulation scheme, we propose a sequence
permutation mechanism to simulate certain quantum states and a generalized
gate array model to simulate quantum algorithm, such as Shor's algorithm and
Grover's algorithm. The research on classical simulation of quantum states
is important, for it not only enables potential beyond quantum computation,
but also provides useful insights into fundamental concepts of quantum
mechanics.
\end{abstract}

\date{today}
\keywords{Quantum computation, Classical entanglement}
\startpage{1}
\email{jianfu@zju.edu.cn}
\maketitle

\section*{Introduction \label{sec1}}

It has been widely known that fast algorithm is not available for many
famous computing problems according to classical computational complexity
theory \cite{Garey}. Quantum computation, however, enormously promotes
computational efficiency with accumulation of computational complexity no
faster than polynomial rate under linear increasement of input size by using
several basic and purely physical features of quantum mechanics, such as
coherent superposition, parallelism, entanglement, measurement collapse etc. 
\cite{Nielsen}. The acceleratory ability of quantum computation is related
to tensor product and quantum entanglement, of which the latter one is
essential to realize quantum computation and quantum communication \cite%
{Jozsa,Braunstein}. Yet quantum algorithm is difficult to be realized for
restrict of quantum system controllability, decoherence property and
measurement randomness \cite{Knill,Nielsen2,Browne,Kok}.

Recently, several researches have proposed a new concept of realization of
classical entanglement based on classical optical fields by introducing a
new freedom degree to realize tensor product in quantum entanglement \cite%
{Toppel}. Further, Ref. \cite{Fu1} proposed that phase modulation by
orthogonal pseudo-random sequence is able to simulate quantum entanglement
effectively. In this scheme, these classical fields with an increased
freedom degree not only realized tensor product structure but also simulated
the nonlocal property of quantum entanglement by using property of
orthogonal pseudo-random sequence like orthogonality, balance and closure.
It is interesting that this simulation creates a Hilbert space of $n2^{n}$
dimensions which is larger than quantum mechanics. Different from quantum
measurement involving collapse, these classical fields are measured in
intensity by optical detectors directly.

The classical simulation of quantum systems, especially of quantum
entanglement has been under investigation for a long time\ \cite%
{Cerf,Massar,Spreeuw}. In addition to potential practical applications in
quantum computation, research on classical simulation systems can help
understand some fundamental concepts in quantum mechanics. However, it has
been pointed out by several authors that classical simulation of quantum
systems exhibits exponentially scaling of physical resources with the number
of quantum particles \cite{Jozsa,Spreeuw}. In Ref. \cite{Spreeuw}, an
optical analogy of quantum systems is introduced, in which the number of
light beams and optical components required grows exponentially with the
number of qubits. In Ref. \cite{Vidal}, a classical protocol to efficiently
simulate any pure-state quantum computation is presented, yet the amount of
entanglement involved is restricted. In Ref. \cite{Jozsa}, it is elucidated
that in classical theory, the state space of a composite system is the
Cartesian product of subsystems, whereas in quantum theory it is the tensor
product. This essential distinction between Cartesian and tensor products is
precisely the phenomenon of quantum entanglement, and viewed as the origin
of the limitation of classical simulation of quantum systems. Therefore it
is of great significance in the classical simulations to realize tensor
product \cite{Toppel,Fu1}.

In wireless and optical communications, orthogonal pseudorandom sequences
have been widely applied to Code Division Multiple Access (CDMA)
communication technology as a way to distinguish different users \cite%
{Viterbi,Peterson}. A set of pseudorandom sequences is generated from a
shift register guided by a Galois field GF($p$), which satisfies orthogonal,
closure and balance properties \cite{Peterson}. In Phase Shift Keying (PSK)
communication systems \cite{PSK}, pseudorandom sequences are used to
modulate the phase of the electromagnetic/optical wave, where a pseudorandom
sequence is mapped to a pseudorandom phase sequence (PPS) values in $\left\{
0,2\pi /p,\cdots ,2\pi \left( p-1\right) /p\right\} $. Guaranteed by the
orthogonal property of the PPS, different electromagnetic/optical waves can
transmit in one communication channel simultaneously without crosstalk, and
can be easily distinguished by implementing a quadrature demodulation
measurement \cite{Viterbi}.

In this paper, we propose a new mechanism based on circular demodulation to
realize simulation of certain quantum state represented by these classical
optical fields. Besides, classical simulation of some other typical quantum
states is discussed, including product state, Bell states, GHZ and W states.
Furthermore, we propose a new scheme to simulate quantum computing by
constructing certain quantum states based on an array of several mode
control gates. Finally, we use this method to simulate Shor's algorithm \cite%
{Shor,Shor2} and Grover's algorithm \cite{Grover}.

The paper is organized as follows: In Section \ref{sec2}, we introduce some
preparing knowledge needed later in this paper. In Section \ref{sec3}, the
existence of the tensor product structure in our simulation is demonstrated
and the classical simulation of several typical quantum states is analyzed.
In Section \ref{sec4}, a generalized gate array model to simulate quantum
computation is proposed. Finally, we summarize our conclusions in Section %
\ref{sec5}.

\section{Preparing Knowledge \label{sec2}}

In this section, we introduce some notation and basic results needed later
in this paper. We first introduce pseudorandom sequences and their
properties. Then we discuss the similarities between classical optical
fields and single-particle quantum states. Finally, we introduce the scheme
of modulation and demodulation on classical optical fields with PPSs.

\subsection{Pseudorandom sequences and their properties \label{sec2.1}}

As far as we know, orthogonal pseudorandom sequences have been widely
applied to CDMA communication technology as a way to distinguish different
users \cite{Viterbi,Peterson}. A set of pseudorandom sequences is generated
from a shift register guided by a Galois Field GF($p$), which satisfies
orthogonal, closure and balance properties. The orthogonal property ensures
that sequences of the set are independent and distinguished each other with
an excellent correlation property. The closure property ensures that any
linear combination of the sequences remains in the same set. The balance
property ensures that the occurrence rate of all non-zero-element in every
sequence is equal, and the the number of zero-elements is exactly one less
than the other elements.

One famous generator of pseudorandom sequences is Linear Feedback Shift
Register (LFSR), which can produce a maximal period sequence, called
m-sequence \cite{Peterson}. We consider an m-sequence of period $N-1$ ($%
N=p^{s}$) generated by a primitive polynomial of degree $s$ over GF($p$).
Since the correlation between different shifts of an m-sequence is almost
zero, they can be used as different codes with their excellent correlation
property. In this regard, the set of $N-1$ m-sequences of length $N-1$ can
be obtained by cyclic shifting of a single m-sequence.

In this paper, we employ pseudorandom phase sequences (PPSs) with $2$-ary
phase shift modulation, which is a well-known modulation format in wireless
and optical communications, including PSK \cite{PSK}. Different from Ref. 
\cite{Fu1}, we choose GF($2$) instead of GF($4$) because the correlation
measurements for Bell inequality are not discussed in this paper. We first
propose a scheme to generate a PPS set $\Xi =\left\{ \lambda ^{\left(
0\right) },\lambda ^{\left( 1\right) },\ldots ,\lambda ^{\left( N-1\right)
}\right\} $ over GF($2$) \cite{Peterson}. $\lambda ^{\left( 0\right) }$ is
an all-$0$ sequence and other sequences can be generated by using the method
as follows:

(1) given a primitive polynomial of degree $s$ over GF($2$), a base sequence
of a length $2^{s}-1$ is generated by using LFSR;

(2) other sequences are obtained by cyclic shifting of the base sequence;

(3) by adding a zero-element to the end of each sequence, the occurrence
rates of all elements in all sequences are equal with each other;

(4) mapping the elements of the sequences to $\{0,\pi /2\}$: $0$ mapping $0$%
, $1$ mapping $\pi /2$.

\begin{figure}[htbp]
\centering\includegraphics[width=5.2736in]{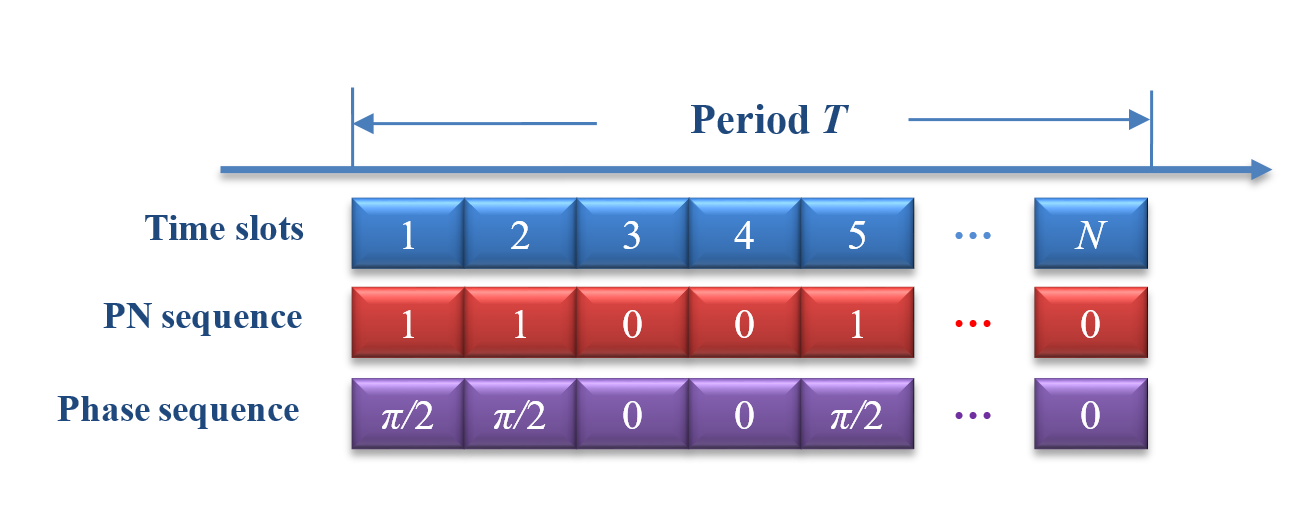}
\caption{The time sequence relationship of the PPS is shown.}
\label{1}
\end{figure}

Further, we define a map $f:\lambda \rightarrow e^{i\lambda }$ on the set of 
$\Xi $, and obtain a new sequence set $\Omega =\left\{ \varphi ^{\left(
j\right) }\left\vert \varphi ^{\left( j\right) }=e^{i\lambda ^{\left(
j\right) }},\right. j=0,\ldots ,N-1\right\} $. In Fig. \ref{1}, we
demonstrate the relationship between time slots, an m-sequence and phase
sequence $\lambda ^{\left( i\right) }$ with $N$ phase units $\lambda
^{\left( i\right) }=[\lambda _{1}^{\left( i\right) },\lambda _{2}^{\left(
i\right) },\cdots \lambda _{N}^{\left( i\right) }]$. For better
understanding our scheme, the PPSs in the cases of modulating $7$ classical
optical fields is illustrated below. Using the method mentioned in section %
\ref{sec2}, an m-sequence of length $2^{3}-1$\ is generated by a primitive
polynomial of the lowest degree over $GF(2)$, which is $\left[ 
\begin{array}{ccccccc}
1 & 1 & 1 & 0 & 0 & 1 & 0%
\end{array}%
\right] $. Then we obtain a group that includes $8$ PPSs of length $8$: $%
\left\{ \lambda ^{\left( 0\right) },\ldots ,\lambda ^{\left( 7\right)
}\right\} $, of which in exception to $\lambda ^{\left( 0\right) }$, all
PPSs can be used to modulate classical optical fields to simulate quantum
states of up to $7$ particles expect $\lambda ^{\left( 0\right) }$, for
example,%
\begin{eqnarray}
\lambda ^{\left( 1\right) } &=&\left[ 
\begin{array}{cccccccc}
1 & 1 & 1 & 0 & 0 & 1 & 0 & 0%
\end{array}%
\right] \times \pi /2, \\
\lambda ^{\left( 2\right) } &=&\left[ 
\begin{array}{cccccccc}
1 & 1 & 0 & 0 & 1 & 0 & 1 & 0%
\end{array}%
\right] \times \pi /2,  \nonumber \\
\lambda ^{\left( 3\right) } &=&\left[ 
\begin{array}{cccccccc}
1 & 0 & 0 & 1 & 0 & 1 & 1 & 0%
\end{array}%
\right] \times \pi /2,  \nonumber \\
\lambda ^{\left( 4\right) } &=&\left[ 
\begin{array}{cccccccc}
0 & 0 & 1 & 0 & 1 & 1 & 1 & 0%
\end{array}%
\right] \times \pi /2,  \nonumber \\
\lambda ^{\left( 5\right) } &=&\left[ 
\begin{array}{cccccccc}
0 & 1 & 0 & 1 & 1 & 1 & 0 & 0%
\end{array}%
\right] \times \pi /2,  \nonumber \\
\lambda ^{\left( 6\right) } &=&\left[ 
\begin{array}{cccccccc}
1 & 0 & 1 & 1 & 1 & 0 & 0 & 0%
\end{array}%
\right] \times \pi /2,  \nonumber \\
\lambda ^{\left( 7\right) } &=&\left[ 
\begin{array}{cccccccc}
0 & 1 & 1 & 1 & 0 & 0 & 1 & 0%
\end{array}%
\right] \times \pi /2.  \nonumber
\end{eqnarray}

According to the properties of m-sequence, we can obtain following
properties of the set $\Omega $, (1) the closure property: the product of
any sequences remains in the same set; (2) the balance property: in
exception to $\varphi ^{\left( 0\right) }$, any sequences of the set $\Omega 
$ satisfy 
\begin{equation}
\sum\limits_{k=1}^{N}e^{i\theta }\varphi _{k}^{\left( j\right)
}=\sum\limits_{k=1}^{N}e^{i\left( \lambda _{k}^{\left( j\right) }+\theta
\right) }=0,\forall \theta \in 
%TCIMACRO{\U{211d} }%
%BeginExpansion
\mathbb{R}
%EndExpansion
;  \label{eq0}
\end{equation}%
(3) the orthogonal property: any two sequences satisfy the following
normalized correlation 
\begin{eqnarray}
E\left( {\varphi ^{\left( i\right) },\varphi ^{\left( j\right) }}\right) &=&%
\frac{1}{N}\sum\limits_{k=1}^{N}{\varphi _{k}^{\left( i\right) }\varphi
_{k}^{\left( j\right) \ast }}  \label{eq1} \\
&=&\left\{ 
\begin{array}{cc}
1, & i=j \\ 
0, & i\neq j%
\end{array}%
\right. .  \nonumber
\end{eqnarray}

In conclusion, the map $f$\ corresponds to the modulation of PPSs of $\Omega 
$ on classical optical fields. According to the properties above, the
classical optical fields modulated with different PPSs become independent
and distinguishable.

\subsection{Similarities between classical optical field and single-particle
quantum states \label{sec2.2}}

We note the similarities between Maxwell equation and Schr\"{o}dinger
equation. In fact, some properties of quantum information are wave
properties, where the wave is not necessary to be a quantum wave \cite%
{Spreeuw}. Analogous to quantum states, classical optical fields also obey a
superposition principle, and can be transformed to any superposition state
by unitary transformations. Those analogous properties made the simulation
of quantum states using\textbf{\ }polarization or transverse modes of
classical optical fields possible \cite{Fu,dragoman,Lee}\textbf{. }

We first consider two orthogonal modes (polarization or transverse), as the
classical simulation of quantum bits (qubits) $\left\vert 0\right\rangle $\
and\ $\left\vert 1\right\rangle $\ \cite{Spreeuw,Fu}. 
\begin{equation}
\left\vert 0\right\rangle =\left( 
\begin{array}{c}
1 \\ 
0%
\end{array}%
\right) ,\left\vert 1\right\rangle =\left( 
\begin{array}{c}
0 \\ 
1%
\end{array}%
\right) .  \label{eq2}
\end{equation}%
Thus, any quantum state of a single particle\ can be simulated by a
corresponding classical mode superposition field, as follows 
\begin{equation}
\left\vert \psi \right\rangle =\alpha \left\vert 0\right\rangle +\beta
\left\vert 1\right\rangle ,\left\vert \alpha \right\vert ^{2}+\left\vert
\beta \right\vert ^{2}=1,\left( \alpha ,\beta \in 
%TCIMACRO{\U{2102} }%
%BeginExpansion
\mathbb{C}
%EndExpansion
\right) .  \label{eq3}
\end{equation}%
Obviously, all the mode superposition fields can span a Hilbert space, where
we can perform unitary transformations to transform the mode state. For
example, the unitary transformation $U\left( \chi ,\theta \right) $ is
defined 
\begin{equation}
U\left( \chi ,\theta \right) =e^{i\chi \left( \sigma _{x}\cos \theta +\sigma
_{y}\sin \theta \right) }=\left( 
\begin{array}{cc}
\cos \chi & -ie^{i\theta }\sin \chi \\ 
ie^{-i\theta }\sin \chi & \cos \chi%
\end{array}%
\right) ,  \label{eq4}
\end{equation}%
where $\sigma _{x}$, $\sigma _{y}$ are Pauli matrices. The modes $\left\vert
0\right\rangle $ and $\left\vert 1\right\rangle $ can be transformed to mode
superposition by using $U\left( \chi ,\theta \right) $, respectively, as
follows%
\begin{eqnarray}
U\left( \chi ,\theta \right) \left\vert 0\right\rangle &=&\cos \chi
\left\vert 0\right\rangle +ie^{i\theta }\sin \chi \left\vert 1\right\rangle ,
\label{eq6} \\
U\left( \chi ,\theta \right) \left\vert 1\right\rangle &=&\cos \chi
\left\vert 1\right\rangle -ie^{-i\theta }\sin \chi \left\vert 0\right\rangle
.  \nonumber
\end{eqnarray}

Now, we consider some devices with one input and two outputs, such as beam
or mode splitters, which split one input field $\left\vert \psi
_{in}\right\rangle =\alpha \left\vert 0\right\rangle +\beta \left\vert
1\right\rangle $ into two output fields $\left\vert \psi _{out}^{\left(
a\right) }\right\rangle $ and $\left\vert \psi _{out}^{\left( b\right)
}\right\rangle $. For the case of beam splitters, the output fields are $%
\left\vert \psi _{out}^{\left( a\right) }\right\rangle =C_{a}\left( \alpha
\left\vert 0\right\rangle +\beta e^{i\phi _{a}}\left\vert 1\right\rangle
\right) $ and $\left\vert \psi _{out}^{\left( b\right) }\right\rangle
=C_{b}\left( \alpha \left\vert 0\right\rangle +\beta e^{i\phi
_{b}}\left\vert 1\right\rangle \right) $ with an arbitrary power ratio $%
\left\vert C_{a}\right\vert ^{2}:\left\vert C_{b}\right\vert ^{2}$ between
the output beams, where $\phi _{a,b}$ are the additional phases due to the
splitter. For the case of mode splitters, the output fields are $\left\vert
\psi _{out}^{\left( a\right) }\right\rangle =\alpha e^{i\phi _{a}}\left\vert
0\right\rangle $ and $\left\vert \psi _{out}^{\left( b\right) }\right\rangle
=\beta e^{i\phi _{b}}\left\vert 1\right\rangle $, where $\phi _{a,b}$ are
also the additional phases. Conversely, the devices can act as beam or mode
combiners in which beams or modes from two inputs are combined into one
output.

\subsection{Modulation and demodulation on classical optical fields with
pseudorandom phase sequences \label{sec2.3}}

\begin{figure}[htbp]
\centering\includegraphics[width=2.8859in]{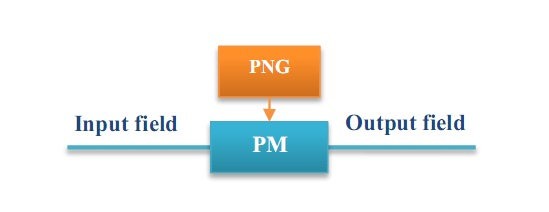}
\caption{The PPS encoding scheme for one input field is shown, where PNG
denotes the pseudorandom number generator and PM denotes the phase
modulator. }
\label{2}
\end{figure}

We first consider the modulation process on a classical optical field with a
PPS. Similar to PSK system, choosing a PPS $\lambda ^{\left( i\right) }$ in
the set of $\Xi $, the phase of the field can be modulated by a phase
modulator (PM) that controlled by a pseudorandom number generator (PNG), the
scheme is shown in Fig. \ref{2}. If the input is a single-mode field, it can
be transformed to mode superposition by performing a unitary transformation
after the modulation.

\begin{figure}[htbp]
\centering\includegraphics[width=4.1891in]{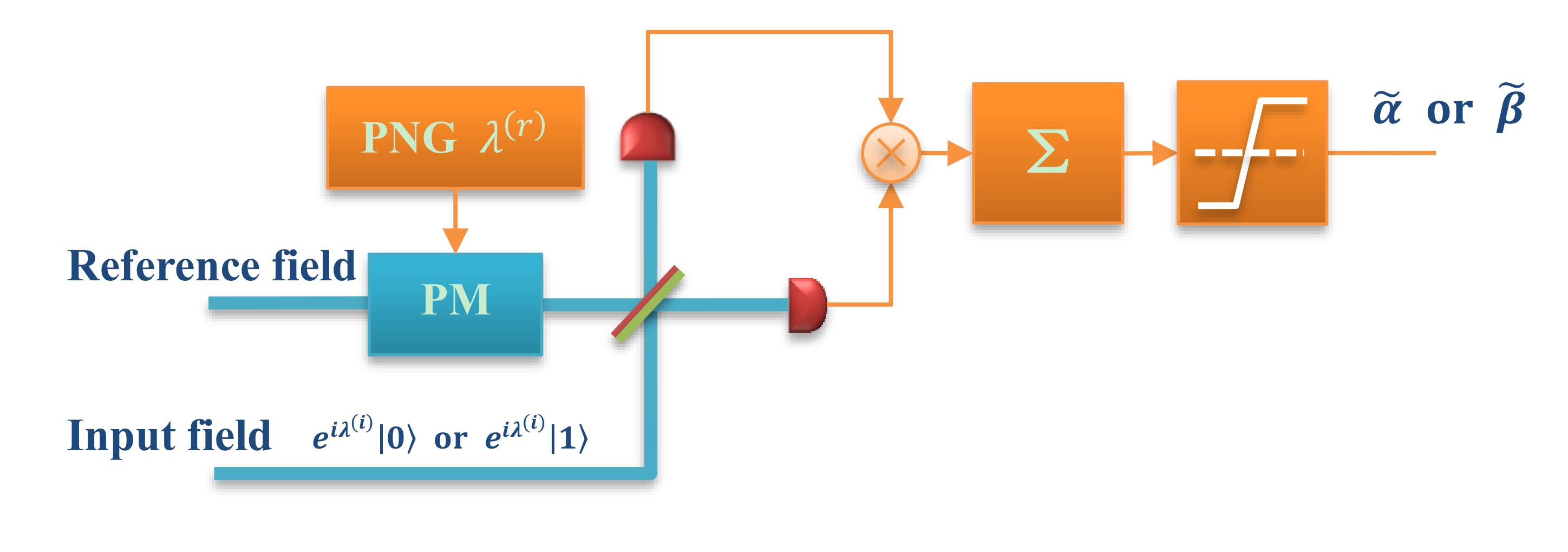}
\caption{The PPS quadrature demodulation scheme for one input field with
single mode $e^{i\protect\lambda ^{\left( i\right) }}\left\vert
0\right\rangle $ or $e^{i\protect\lambda ^{\left( i\right) }}\left\vert
1\right\rangle $ is shown.}
\label{3}
\end{figure}

\begin{figure}[htbp]
\centering\includegraphics[width=3.813in]{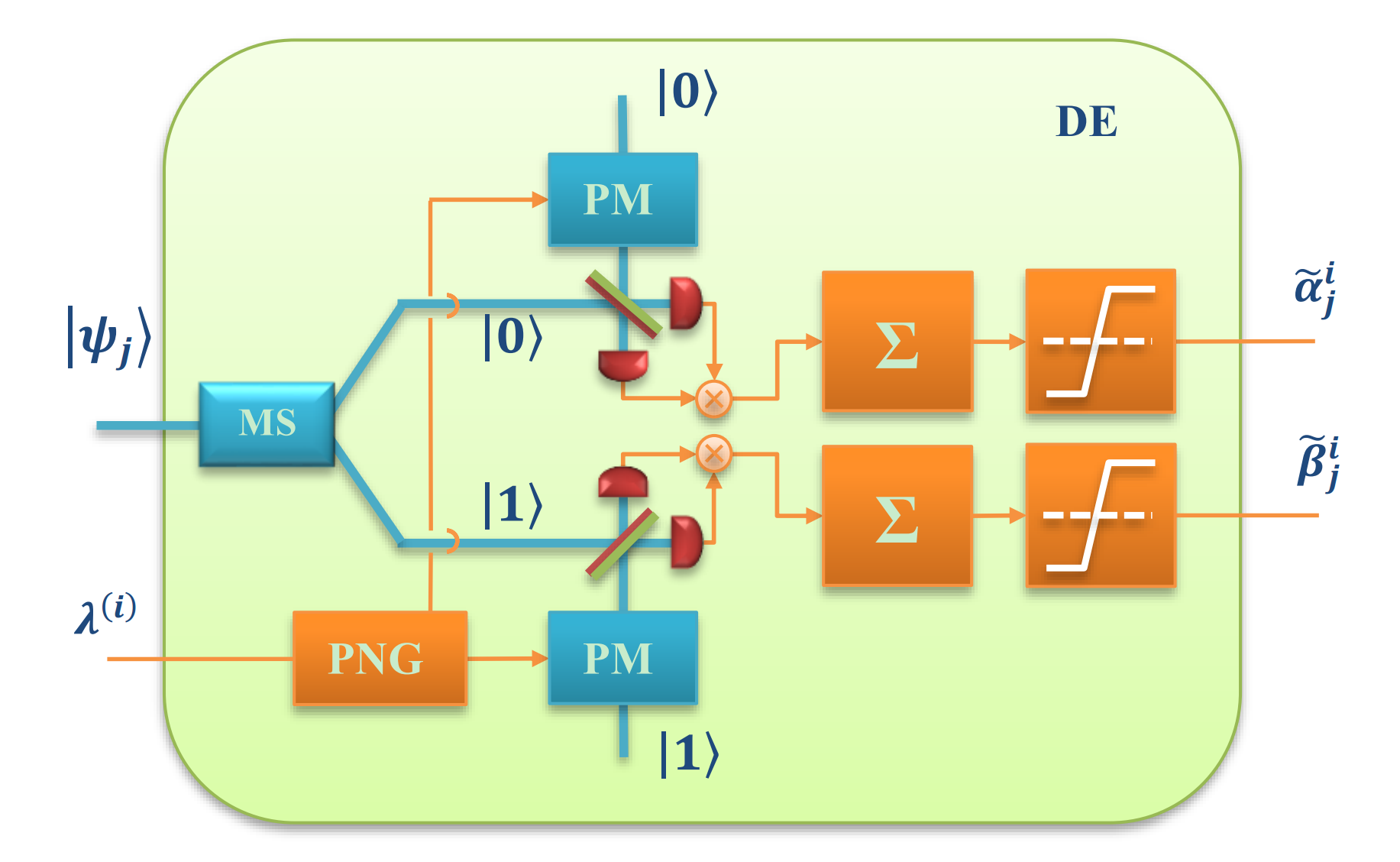}
\caption{The PPS quadrature demodulation scheme for one field with two
orthogonal modes is shown, where the blue block MS denote the mode splitter.}
\label{4}
\end{figure}

Then we consider the quadrature demodulation process of a modulated
classical optical field. Quadrature demodulation is a coherent detection
process that allows the simultaneous measurement of conjugate quadrature
components via homodyning the emerging beams with the input and reference
fields by using a balanced beam splitting, where the reference field is
modulated with a PPS $\lambda ^{\left( r\right) }$. The differenced signals
of two output detectors are then summed over $N$ and sampled to yield the
decision variable. We can express the demodulation process in mathematical
form 
\begin{equation}
I\left( \lambda ^{\left( i\right) },\lambda ^{\left( r\right) }\right) =%
\frac{1}{N}\sum_{k=1}^{N}\cos \left( \lambda _{k}^{\left( i\right) }-\lambda
_{k}^{\left( r\right) }\right) =\left\{ 
\begin{array}{cc}
1, & i=r \\ 
0, & i\neq r%
\end{array}%
\right. ,  \label{eq11}
\end{equation}%
where $\lambda _{k}^{\left( i\right) },\lambda _{k}^{\left( r\right) }$ are
the PPSs of the input and reference fields respectively. The output decision
variable is $1$\ if and only if $\lambda _{k}^{\left( i\right) },\lambda
_{k}^{\left( r\right) }$\ are equal; otherwise the output decision variable
is $0$. The results are guaranteed by the properties of PPSs. If the input
is a single-mode field, the scheme shown in Fig. \ref{3} is employed to
perform quadrature demodulation. Otherwise the scheme shown in Fig. \ref{4}
is used, in which the input field $\left\vert \psi _{i}\right\rangle
=e^{i\lambda ^{\left( i\right) }}\left( \alpha _{i}\left\vert 0\right\rangle
+\beta _{i}\left\vert 1\right\rangle \right) $ is first splitted into two
fields $\alpha _{i}e^{i\lambda ^{\left( i\right) }}\left\vert 0\right\rangle 
$ and $\beta _{i}e^{i\lambda ^{\left( i\right) }}\left\vert 1\right\rangle $%
, and two coherent detection processes are then performed on the two fields
respectively. Noteworthily, the modes of the reference fields must be
consistent with the two output fields. Thus there are two output decision
variables $\tilde{\alpha}$\ and $\tilde{\beta}$, which correspond to the
modes $\left\vert 0\right\rangle $\ and $\left\vert 1\right\rangle $,
respectively. We define $\left( \tilde{\alpha},\tilde{\beta}\right) $ as a
mode status that displays not only existence of mode also phases due to the
polarity of signals. Besides the quadrature demodulation, we can also easily
measure the amplitudes of each modes $\left\vert \alpha _{i}\right\vert
,\left\vert \beta _{i}\right\vert $ after mode spitting in the scheme.
However, we will not deal with normalization of superposition coefficients $%
\alpha $\ and $\beta $ in our scheme, because the amplitudes of each modes
are not important.

\section{\textbf{Simulation} of multiparticle quantum states \label{sec3}}

We discuss simulation of multiparticle quantum states using classical
optical fields modulated with PPSs in this section. We first demonstrate
that $n$ classical optical fields modulated with $n$\ different PPSs can
constitute a similar $n2^{n}$-dimensional Hilbert space that contains a
tensor product structure \cite{Fu1}. Besides, by performing quadrature
demodulation scheme, we can obtain the mode status matrix of the simulating
classical optical fields, based on which we propose a sequence permutation
mechanism to simulate the quantum states\textbf{. }Classical simulations of
some typical quantum states are then discussed, including product state,
Bell states, GHZ state and W state.

\subsection{Classical optical fields modulated with pseudorandom phase
sequences and their tensor product structure \label{sec3.1}}

For convenience, here we consider two classical optical fields modulated
with PPSs and their tensor product structure. Chosen any two PPSs of $%
\lambda ^{\left( a\right) }$ and $\lambda ^{\left( b\right) }$ from the set $%
\Xi $, any two fields modulated with the PPSs can be expressed as follows, 
\begin{eqnarray}
\left\vert \psi _{a}\right\rangle &=&e^{i\lambda ^{\left( a\right) }}\left(
\alpha _{a}\left\vert 0\right\rangle +\beta _{a}\left\vert 1\right\rangle
\right) ,  \label{eq13} \\
\left\vert \psi _{b}\right\rangle &=&e^{i\lambda ^{\left( b\right) }}\left(
\alpha _{b}\left\vert 0\right\rangle +\beta _{b}\left\vert 1\right\rangle
\right) .  \nonumber
\end{eqnarray}%
We define the inner product of two fields $\left\vert \psi _{a}\right\rangle 
$ and $\left\vert \psi _{b}\right\rangle $ as follows,%
\begin{equation}
\left\langle \psi _{a}|\psi _{b}\right\rangle =\frac{1}{N}%
\sum\limits_{k=1}^{N}e^{i\left( \lambda _{k}^{\left( b\right) }-\lambda
_{k}^{\left( a\right) }\right) }\left( \alpha _{b}\alpha _{a}^{\ast }+\beta
_{b}\beta _{a}^{\ast }\right) .  \label{eq14}
\end{equation}%
According to the properties of the PPSs, we can easily obtain%
\begin{equation}
\left\langle \psi _{a}|\psi _{b}\right\rangle =\left\{ 
\begin{array}{cc}
1, & a=b \\ 
0, & a\neq b%
\end{array}%
\right. ,  \label{eq15}
\end{equation}%
which shows that two fields modulated with two different PPSs are
orthogonal. The orthogonal property make contribution to the tensor product
structure of multiple fields. The direct product states of the two fields
can be expressed as follows,%
\begin{equation}
\left\vert \psi _{a}\right\rangle \otimes \left\vert \psi _{b}\right\rangle
=e^{i\left( \lambda ^{\left( a\right) }+\lambda ^{\left( b\right) }\right)
}\left( \alpha _{a}\alpha _{b}\left\vert 0\right\rangle \left\vert
0\right\rangle +\alpha _{a}\beta _{b}\left\vert 0\right\rangle \left\vert
1\right\rangle +\beta _{a}\alpha _{b}\left\vert 1\right\rangle \left\vert
0\right\rangle +\beta _{a}\beta _{b}\left\vert 1\right\rangle \left\vert
1\right\rangle \right) ,  \label{eq16}
\end{equation}%
where $\lambda ^{\left( a\right) }+\lambda ^{\left( b\right) }$ remains in
the set $\Xi $ due to the closure property.

Quantum entanglement is only defined for Hilbert spaces that have a rigorous
tensor product structure in terms of subsystems. As shown in Ref. \cite{Fu1}%
, for any $n$ classical optical fields with $n$ PPSs, the Basis for Hilbert
space of simulation is spanned by $\left\{ e^{i\lambda ^{(j)}}\left\vert
i_{1}i_{2}\ldots i_{n}\right\rangle |j=1\ldots n,i_{k}=0\;or\;1\right\} $,
with a total base state number of $n2^{n}$. It is obvious that the Hilbert
simulation space is greater than what is required for simulation of quantum
state. The tensor product structure and the efficient classical simulation
of quantum entanglement have been discussed. Here we construct a similar
structure of multiple classical optical fields based on the efficient
classical simulation of quantum entanglement.

\subsection{\textbf{Simulation} of quantum states based on classical optical
fields \label{sec3.2}}

We have discussed quadrature demodulation process in Sec. \ref{sec2.3}. Here
we discuss how to simulate quantum state based on classical optical fields
with the help of quadrature demodulation.

First, we consider the general form of $n$ classical optical fields
modulated with PPSs $\left\{ \lambda ^{\left( 1\right) },\ldots ,\lambda
^{\left( n\right) }\right\} $ chosen from the set $\Xi $, and the states can
be expressed as follows, 
\begin{equation}
\left\vert \psi _{k}\right\rangle =\sum_{i=1}^{n}\alpha
_{k}^{(i)}e^{i\lambda ^{(i)}}\left\vert 0\right\rangle +\sum_{j=1}^{n}\beta
_{k}^{(j)}e^{i\lambda ^{(j)}}\left\vert 1\right\rangle .  \label{eq17}
\end{equation}%
It is noteworthy that although multiple PPSs are superimposed on both modes
of the fields, all of the PPSs can be demodulated and discriminated by
performing the quadrature demodulation introduced in Sec. \ref{sec2.3},
which has already been verified by many actual communication systems \cite%
{Viterbi,Peterson,PSK}.

\begin{figure}[htbp]
\centering\includegraphics[width=5.2736in]{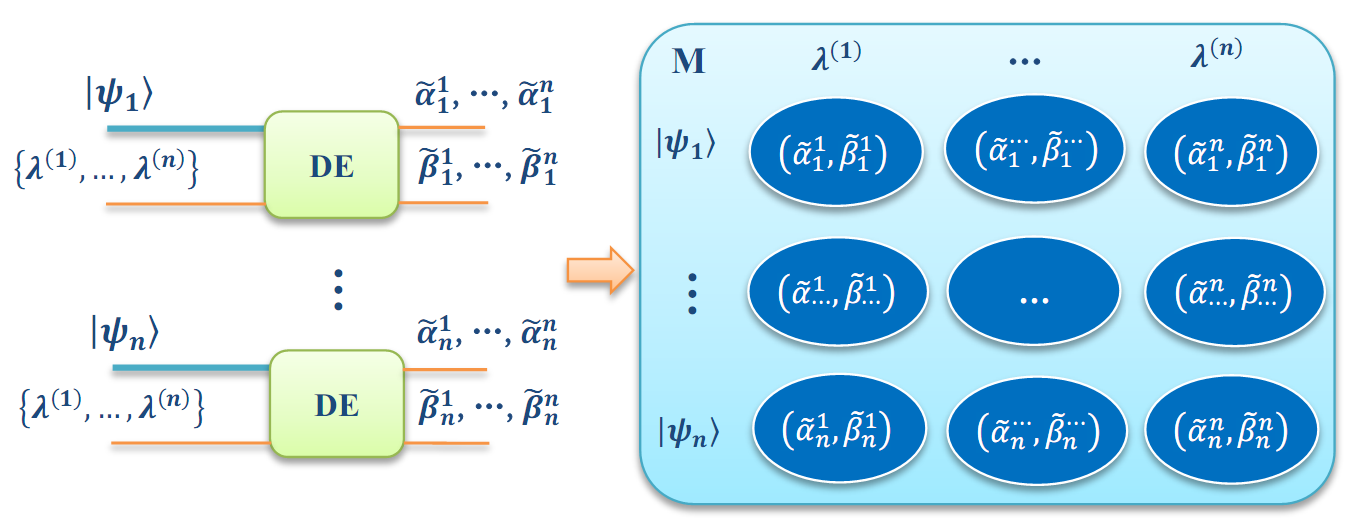}
\caption{The PPS quadrature demodulation scheme for multiple input fields is
shown, where the DE block is shown in Fig. \protect\ref{4}.}
\label{5}
\end{figure}

Now we propose a scheme, as shown in Fig. \ref{4}, to perform the quadrature
demodulation introduced in Sec. \ref{sec2.3}. In the scheme, quadrature
demodulations are performed on each field, in which the reference PPSs are
ergodic on $\left\{ \lambda ^{\left( 1\right) },\ldots ,\lambda ^{\left(
n\right) }\right\} $. Thus a mode status matrix $M\left( \tilde{\alpha}%
_{i}^{j},\tilde{\beta}_{i}^{j}\right) $, as shown in Fig. \ref{5}, can be
obtained by performing $n$ quadrature demodulations on the $n$ classical
optical fields. Of the matrix $M\left( \tilde{\alpha}_{i}^{j},\tilde{\beta}%
_{i}^{j}\right) $, each element is the mode status of the $i$th classical
optical field when the reference PPS is $\lambda ^{\left( j\right) }$. The
element takes one of four possible values: $\left( 1,0\right) ,\left(
0,1\right) ,\left( 1,1\right) $ or $0$, denote that the PPS $\lambda
^{\left( j\right) }$ is modulated on mode $\left\vert 0\right\rangle $ of
the $i$th classical optical field, on mode $\left\vert 1\right\rangle $, on
both $\left\vert 0\right\rangle $ and $\left\vert 1\right\rangle $, on
neither $\left\vert 0\right\rangle $ nor $\left\vert 1\right\rangle $,
respectively. It is noteworthy that different modulation of the $n$
classical optical fields correspond to different mode status matrixes, and
vice versa. Thus we obtain a one-to-one correspondence relationship between
the $n$ classical optical fields and the mode status matrix. Besides,
further discussion will show that structure of quantum states and quantum
entanglement can be revealed in the mode status matrix, which means that a
correspondence can also be obtained between the mode status matrix and
quantum states. Thus we treat the mode status matrix as a bridge to connect
the simulating fields and the quantum states.

\begin{figure}[htbp]
\centering\includegraphics[width=4.804in]{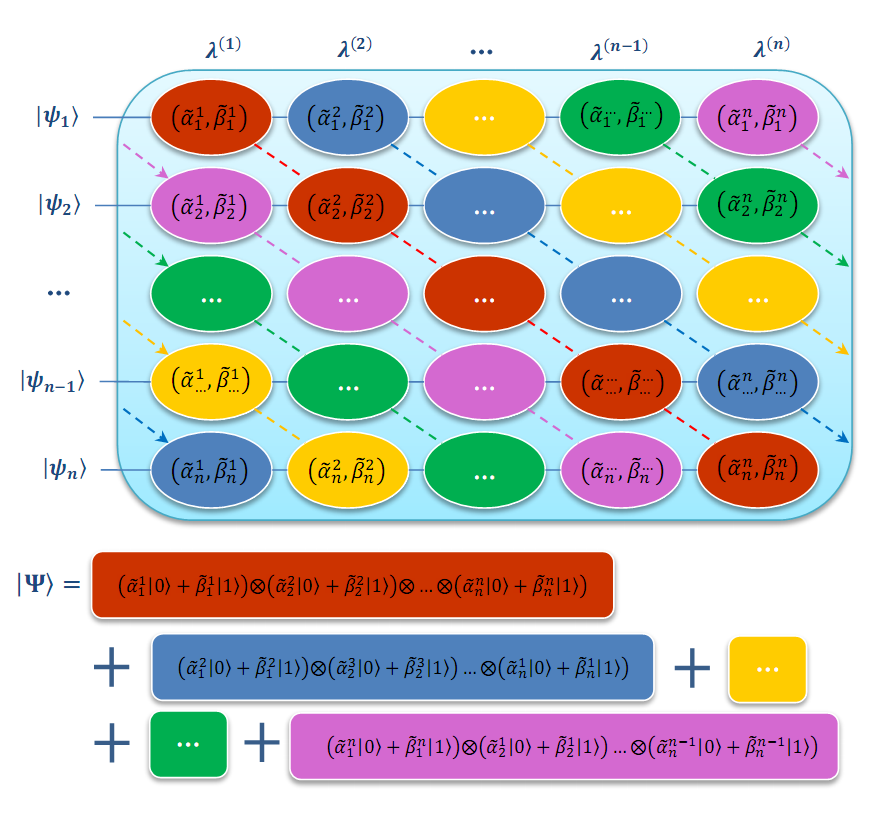}
\caption{The scheme to simulate quantum state is shown, the mode status
matrix $M\left( \tilde{\protect\alpha}_{i}^{j},\tilde{\protect\beta}%
_{i}^{j}\right) $ related to $i$th classical optical field and the reference
PPS $\protect\lambda ^{\left( j\right) }$, in which the mode status with
same color for same sequence permutation.}
\label{6}
\end{figure}

Now we consider how to construct the states based on $M$ matrix,
respectively. Thus we propose a sequence permutation mechanism to simulate
each $\left\vert \Psi \right\rangle $ based on $M$ matrix, which is one of
the simplest mechanisms for sequence ergodic ensemble. Assumed $\left\vert
\Psi \right\rangle $ contains $n$ classical optical fields with $n$ PPSs,
namely the corresponding matrix contains $n$ rows and $n$ columns, the
sequence permutation is arranged as 
\begin{equation}
R_{1}=\left\{ \lambda ^{\left( 1\right) },\lambda ^{\left( 2\right) },\ldots
,\lambda ^{\left( n\right) }\right\} ,R_{2}=\left\{ \lambda ^{\left(
2\right) },\lambda ^{\left( 3\right) },\ldots ,\lambda ^{\left( 1\right)
}\right\} ,\ldots ,R_{n}=\left\{ \lambda ^{\left( n\right) },\lambda
^{\left( 1\right) },\ldots ,\lambda ^{\left( n-1\right) }\right\} .
\label{eq22}
\end{equation}%
As shown in Fig. \ref{6}, we can obtain the mode status with same color for
same sequence permutation, such as the red color corresponding to $R_{1}$,
the blue color corresponding to $R_{2}$, etc. We obtain a direct product of $%
n$ items for each $R_{r}$, and the simulated quantum state is the
superposition of the $n$ product items. Therefore we obtain%
\begin{eqnarray}
\left\vert \Psi \right\rangle &=&\left( \tilde{\alpha}_{1}^{1}\left\vert
0\right\rangle +\tilde{\beta}_{1}^{1}\left\vert 1\right\rangle \right)
\otimes \ldots \otimes \left( \tilde{\alpha}_{n}^{n}\left\vert
0\right\rangle +\tilde{\beta}_{n}^{n}\left\vert 1\right\rangle \right)
+\left( \tilde{\alpha}_{1}^{2}\left\vert 0\right\rangle +\tilde{\beta}%
_{1}^{2}\left\vert 1\right\rangle \right) \otimes \ldots \otimes \left( 
\tilde{\alpha}_{n}^{1}\left\vert 0\right\rangle +\tilde{\beta}%
_{n}^{1}\left\vert 1\right\rangle \right)  \nonumber \\
&&+\ldots +\left( \tilde{\alpha}_{1}^{n}\left\vert 0\right\rangle +\tilde{%
\beta}_{1}^{n}\left\vert 1\right\rangle \right) \otimes \ldots \otimes
\left( \tilde{\alpha}_{n}^{n-1}\left\vert 0\right\rangle +\tilde{\beta}%
_{n}^{n-1}\left\vert 1\right\rangle \right) ,  \label{eq23}
\end{eqnarray}%
where $\left( \tilde{\alpha}_{i}^{j},\tilde{\beta}_{i}^{j}\right) $ is the
mode status obtained from the matrix $M\left( \tilde{\alpha}_{i}^{j},\tilde{%
\beta}_{i}^{j}\right) $.

It is noteworthy that the mechanism we proposed above is one of the feasible
ways to simulate the quantum state based on the mode status matrix. Other
mechanisms may also work, as long as a sequence ergodic ensemble is obtained
in the mechanism. The sequence permutation mechanism above can successfully
simulate many quantum states, including the product states, Bell states, GHZ
states and W states. We will discuss the related contents in next subsection.

\subsection{\textbf{Simulations} of several typical quantum states \label%
{sec3.3}}

In this subsection, we discuss classical simulations of several typical
quantum states, including product state, Bell states, GHZ state and W state.

\subsubsection{Product state \label{sec3.3.1}}

First, we discuss classical simulation of $n$\ quantum product state. The
simulation fields are shown as follows 
\begin{eqnarray}
\left\vert \psi _{1}\right\rangle &=&e^{i\lambda ^{\left( 1\right) }}\left(
\left\vert 0\right\rangle +\left\vert 1\right\rangle \right) ,  \label{eq18}
\\
&&......  \nonumber \\
\left\vert \psi _{n}\right\rangle &=&e^{i\lambda ^{\left( n\right) }}\left(
\left\vert 0\right\rangle +\left\vert 1\right\rangle \right) .  \nonumber
\end{eqnarray}%
By employing the scheme as shown in Fig. \ref{6}, we obtain the mode status
matrix 
\begin{equation}
M\left( \tilde{\alpha}_{i}^{j},\tilde{\beta}_{i}^{j}\right) =\left( 
\begin{array}{ccc}
\left( 1,1\right) &  & 0 \\ 
& \ddots &  \\ 
0 &  & \left( 1,1\right)%
\end{array}%
\right) ,  \label{eq24}
\end{equation}%
which demonstrates that each classical optical field is the superposition of
two orthogonal modes and no entanglement is involved. According to Eq. (\ref%
{eq23}), we obtain 
\begin{eqnarray}
\left\vert \Psi \right\rangle &=&\left( \left\vert 0\right\rangle
+\left\vert 1\right\rangle \right) \otimes \ldots \otimes \left( \left\vert
0\right\rangle +\left\vert 1\right\rangle \right)  \label{eq25} \\
&=&\left\vert 0\ldots 0\right\rangle +\left\vert 0\ldots 1\right\rangle
+\ldots +\left\vert 1\ldots 1\right\rangle ,  \nonumber
\end{eqnarray}%
where $\left\vert i_{1}\ldots i_{n}\right\rangle \equiv \left\vert
i_{1}\right\rangle \otimes \ldots \otimes \left\vert i_{n}\right\rangle
,\left( i_{n}=0\ or\ 1\right) $, which is same as a quantum product state
expect a normalization factor.

\subsubsection{Bell states \label{sec3.3.2}}

Now we discuss classical simulation of one of the four Bell states $%
\left\vert \Psi ^{+}\right\rangle =\frac{1}{\sqrt{2}}\left( \left\vert
0_{1}\right\rangle \left\vert 0_{2}\right\rangle +\left\vert
1_{1}\right\rangle \left\vert 1_{2}\right\rangle \right) $, which contains
two classical optical fields as follows%
\begin{eqnarray}
\left\vert \psi _{1}\right\rangle &=&e^{i\lambda ^{\left( 1\right)
}}\left\vert 0\right\rangle +e^{i\lambda ^{\left( 2\right) }}\left\vert
1\right\rangle ,  \label{eq26} \\
\left\vert \psi _{2}\right\rangle &=&e^{i\lambda ^{\left( 2\right)
}}\left\vert 0\right\rangle +e^{i\lambda ^{\left( 1\right) }}\left\vert
1\right\rangle .  \nonumber
\end{eqnarray}%
By employing the scheme as shown in Fig. \ref{6}, we obtain the mode status
matrix%
\begin{equation}
M\left( \tilde{\alpha}_{i}^{j},\tilde{\beta}_{i}^{j}\right) =\left( 
\begin{array}{cc}
\left( 1,0\right) & \left( 0,1\right) \\ 
\left( 0,1\right) & \left( 1,0\right)%
\end{array}%
\right) .  \label{eq27}
\end{equation}%
We note that in this case, the mode status matrix is irreducible, which
corresponds to an entanglement state. According to the sequence permutation
mechanism, we obtain that $R_{1}=\{\lambda ^{\left( 1\right) },\lambda
^{\left( 2\right) }\}$ and $R_{2}=\{\lambda ^{\left( 2\right) },\lambda
^{\left( 1\right) }\}$. Based on the mode status matrix, for the selection
of $R_{1}$, we obtain $\left\vert 0\right\rangle \otimes \left\vert
0\right\rangle $; for the selection of $R_{2}$, we obtain $\left\vert
1\right\rangle \otimes \left\vert 1\right\rangle $. If we randomly choose
one selection between $R_{1}$ and $R_{2}$, we can randomly obtain one result
between $\left\vert 0\right\rangle \otimes \left\vert 0\right\rangle $ and $%
\left\vert 1\right\rangle \otimes \left\vert 1\right\rangle $, which is
similar with the case of quantum measurement for the Bell state $\left\vert
\Psi ^{+}\right\rangle $. We can simulate the state except a norm based on
the mode status matrix%
\begin{eqnarray}
\left\vert \Psi ^{+}\right\rangle &=&\left( \tilde{\alpha}_{1}^{1}\left\vert
0\right\rangle +\tilde{\beta}_{1}^{1}\left\vert 1\right\rangle \right)
\otimes \left( \tilde{\alpha}_{2}^{2}\left\vert 0\right\rangle +\tilde{\beta}%
_{2}^{2}\left\vert 1\right\rangle \right) +\left( \tilde{\alpha}%
_{1}^{2}\left\vert 0\right\rangle +\tilde{\beta}_{1}^{2}\left\vert
1\right\rangle \right) \otimes \left( \tilde{\alpha}_{2}^{1}\left\vert
0\right\rangle +\tilde{\beta}_{2}^{1}\left\vert 1\right\rangle \right) 
\nonumber \\
&=&\left\vert 00\right\rangle +\left\vert 11\right\rangle
\end{eqnarray}%
which is same as the Bell state $\left\vert \Psi ^{+}\right\rangle $ expect
a normalization factor.

In quantum mechanics, another Bell state $\left\vert \Phi ^{+}\right\rangle $
can be obtained from $\left\vert \Psi ^{+}\right\rangle $ by performing the
unitary transformation $\sigma _{x}:\left\vert 0\right\rangle
\leftrightarrow \left\vert 1\right\rangle $ on one of the particles. Using
the same method, we perform an unitary transformation on $\left\vert \psi
_{b}\right\rangle $ to flip its modes $\left\vert 0\right\rangle
\leftrightarrow \left\vert 1\right\rangle $. Thus we obtain two classical
optical fields as follows%
\begin{eqnarray}
\left\vert \psi _{1}\right\rangle &=&e^{i\lambda ^{\left( 1\right)
}}\left\vert 0\right\rangle +e^{i\lambda ^{\left( 2\right) }}\left\vert
1\right\rangle ,  \label{eq29} \\
\left\vert \psi _{2}\right\rangle &=&e^{i\lambda ^{\left( 2\right)
}}\left\vert 1\right\rangle +e^{i\lambda ^{\left( 1\right) }}\left\vert
0\right\rangle .  \nonumber
\end{eqnarray}%
By employing the scheme as shown in Fig. \ref{6}, we obtain the mode status
matrix%
\begin{equation}
M\left( \tilde{\alpha}_{i}^{j},\tilde{\beta}_{i}^{j}\right) =\left( 
\begin{array}{cc}
\left( 1,0\right) & \left( 0,1\right) \\ 
\left( 1,0\right) & \left( 0,1\right)%
\end{array}%
\right) .  \label{eq30}
\end{equation}%
According to the sequence permutation mechanism, here we obtain $R_{1}$ and $%
R_{2}$ again. As the mode status matrix is different, for $R_{1}$, the
result turns to be $\left\vert 0\right\rangle \otimes \left\vert
1\right\rangle $; for $R_{2}$, we obtain $\left\vert 1\right\rangle \otimes
\left\vert 0\right\rangle $. If we randomly choose one selection between $%
R_{1}$ and $R_{2}$, we can also randomly obtain one result between $%
\left\vert 0\right\rangle \otimes \left\vert 1\right\rangle $ and $%
\left\vert 0\right\rangle \otimes \left\vert 1\right\rangle $, which is
similar with the case of quantum measurement for the Bell state $\left\vert
\Phi ^{+}\right\rangle $. We can simulate the state 
\begin{eqnarray}
\left\vert \Phi ^{+}\right\rangle &=&\left( \tilde{\alpha}_{1}^{1}\left\vert
0\right\rangle +\tilde{\beta}_{1}^{1}\left\vert 1\right\rangle \right)
\otimes \left( \tilde{\alpha}_{2}^{2}\left\vert 0\right\rangle +\tilde{\beta}%
_{2}^{2}\left\vert 1\right\rangle \right) +\left( \tilde{\alpha}%
_{1}^{2}\left\vert 0\right\rangle +\tilde{\beta}_{1}^{2}\left\vert
1\right\rangle \right) \otimes \left( \tilde{\alpha}_{2}^{1}\left\vert
0\right\rangle +\tilde{\beta}_{2}^{1}\left\vert 1\right\rangle \right) 
\nonumber \\
&=&\left\vert 10\right\rangle +\left\vert 01\right\rangle
\end{eqnarray}%
which is same as the Bell state $\left\vert \Phi ^{+}\right\rangle $ expect
a normalization factor. For other two Bell states $\left\vert \Psi
^{-}\right\rangle $ and $\left\vert \Phi ^{-}\right\rangle $, they can be
obtained from $\left\vert \Psi ^{+}\right\rangle $ and $\left\vert \Phi
^{+}\right\rangle $ by using a $\pi $ phase transformation. We can
distinguish $\left\vert \Psi ^{-}\right\rangle $ and $\left\vert \Phi
^{-}\right\rangle $ from $\left\vert \Psi ^{+}\right\rangle $ and $%
\left\vert \Phi ^{+}\right\rangle $ by using the signal polarity of
quadrature demodulation.

\subsubsection{GHZ state \label{sec3.3.3}}

For tripartite systems there are only two different classes of genuine
tripartite entanglement, the GHZ class and the W class \cite%
{Greenberger,Nielsen}. First we discuss the classical simulation of GHZ
state $\left\vert \Psi _{GHZ}\right\rangle =\frac{1}{\sqrt{2}}\left(
\left\vert 0_{1}\right\rangle \left\vert 0_{2}\right\rangle \left\vert
0_{3}\right\rangle +\left\vert 1_{1}\right\rangle \left\vert
1_{2}\right\rangle \left\vert 1_{3}\right\rangle \right) $, which contains
three classical optical fields as follows%
\begin{eqnarray}
\left\vert \psi _{1}\right\rangle &=&e^{i\lambda ^{\left( 1\right)
}}\left\vert 0\right\rangle +e^{i\lambda ^{\left( 2\right) }}\left\vert
1\right\rangle ,  \label{eq32} \\
\left\vert \psi _{2}\right\rangle &=&e^{i\lambda ^{\left( 2\right)
}}\left\vert 0\right\rangle +e^{i\lambda ^{\left( 3\right) }}\left\vert
1\right\rangle ,  \nonumber \\
\left\vert \psi _{3}\right\rangle &=&e^{i\lambda ^{\left( 3\right)
}}\left\vert 0\right\rangle +e^{i\lambda ^{\left( 1\right) }}\left\vert
1\right\rangle .  \nonumber
\end{eqnarray}%
Performing the scheme as shown in Fig. \ref{6}, we obtain the mode status
matrix%
\begin{equation}
M\left( \tilde{\alpha}_{i}^{j},\tilde{\beta}_{i}^{j}\right) =\left( 
\begin{array}{ccc}
\left( 1,0\right) & \left( 0,1\right) & 0 \\ 
0 & \left( 1,0\right) & \left( 0,1\right) \\ 
\left( 0,1\right) & 0 & \left( 1,0\right)%
\end{array}%
\right) .  \label{eq33}
\end{equation}%
According to the sequence permutation mechanism, we obtain that $%
R_{1}=\{\lambda ^{\left( 1\right) },\lambda ^{\left( 2\right) },\lambda
^{\left( 3\right) }\}$, $R_{2}=\{\lambda ^{\left( 2\right) },\lambda
^{\left( 3\right) },\lambda ^{\left( 1\right) }\}$ and $R_{3}=\{\lambda
^{\left( 3\right) },\lambda ^{\left( 1\right) },\lambda ^{\left( 2\right)
}\} $. Based on the mode status matrix, for the selection of $R_{1}$, we
obtain $\left\vert 0\right\rangle \otimes \left\vert 0\right\rangle \otimes
\left\vert 0\right\rangle $; for the selection of $R_{2}$, we obtain $%
\left\vert 1\right\rangle \otimes \left\vert 1\right\rangle \otimes
\left\vert 1\right\rangle $; for the selection of $R_{3}$, we obtain
nothing. Thus we can simulate the state based on the mode status matrix%
\begin{eqnarray}
\left\vert \Psi _{GHZ}\right\rangle &=&\left( \tilde{\alpha}%
_{1}^{1}\left\vert 0\right\rangle +\tilde{\beta}_{1}^{1}\left\vert
1\right\rangle \right) \otimes \left( \tilde{\alpha}_{2}^{2}\left\vert
0\right\rangle +\tilde{\beta}_{2}^{2}\left\vert 1\right\rangle \right)
\otimes \left( \tilde{\alpha}_{3}^{3}\left\vert 0\right\rangle +\tilde{\beta}%
_{3}^{3}\left\vert 1\right\rangle \right)  \label{eq34} \\
&&+\left( \tilde{\alpha}_{1}^{2}\left\vert 0\right\rangle +\tilde{\beta}%
_{1}^{2}\left\vert 1\right\rangle \right) \otimes \left( \tilde{\alpha}%
_{2}^{3}\left\vert 0\right\rangle +\tilde{\beta}_{2}^{3}\left\vert
1\right\rangle \right) \otimes \left( \tilde{\alpha}_{3}^{1}\left\vert
0\right\rangle +\tilde{\beta}_{3}^{1}\left\vert 1\right\rangle \right) 
\nonumber \\
&=&\left\vert 000\right\rangle +\left\vert 111\right\rangle .  \nonumber
\end{eqnarray}

\subsubsection{W state \label{sec3.3.4}}

Then we discuss the classical simulation of W state,%
\begin{equation}
\left\vert \Psi _{W}\right\rangle =\frac{1}{\sqrt{3}}\left( \left\vert
1_{1}\right\rangle \left\vert 0_{2}\right\rangle \left\vert
0_{3}\right\rangle +\left\vert 0_{1}\right\rangle \left\vert
1_{2}\right\rangle \left\vert 0_{3}\right\rangle +\left\vert
0_{1}\right\rangle \left\vert 0_{2}\right\rangle \left\vert
1_{3}\right\rangle \right) ,  \label{eq35}
\end{equation}%
which contains three classical optical fields as follows 
\begin{eqnarray}
\left\vert \psi _{1}\right\rangle &=&e^{i\lambda ^{\left( 1\right)
}}\left\vert 1\right\rangle +e^{i\lambda ^{\left( 2\right) }}\left\vert
0\right\rangle +e^{i\lambda ^{\left( 3\right) }}\left\vert 0\right\rangle ,
\label{eq36} \\
\left\vert \psi _{2}\right\rangle &=&e^{i\lambda ^{\left( 1\right)
}}\left\vert 1\right\rangle +e^{i\lambda ^{\left( 2\right) }}\left\vert
0\right\rangle +e^{i\lambda ^{\left( 3\right) }}\left\vert 0\right\rangle , 
\nonumber \\
\left\vert \psi _{3}\right\rangle &=&e^{i\lambda ^{\left( 1\right)
}}\left\vert 1\right\rangle +e^{i\lambda ^{\left( 2\right) }}\left\vert
0\right\rangle +e^{i\lambda ^{\left( 3\right) }}\left\vert 0\right\rangle . 
\nonumber
\end{eqnarray}%
It is noteworthy that the three classical optical fields can be produced
from one single field by using two beam splitters, which is quite similar
with the generation of W state in quantum mechanics. Performing the same
scheme, we obtain the mode status matrix%
\begin{equation}
M\left( \tilde{\alpha}_{i}^{j},\tilde{\beta}_{i}^{j}\right) =\left( 
\begin{array}{ccc}
\left( 0,1\right) & \left( 1,0\right) & \left( 1,0\right) \\ 
\left( 0,1\right) & \left( 1,0\right) & \left( 1,0\right) \\ 
\left( 0,1\right) & \left( 1,0\right) & \left( 1,0\right)%
\end{array}%
\right) .  \label{eq37}
\end{equation}%
According to the sequence permutation mechanism, we use $R_{1}$, $R_{2}$ and 
$R_{3}$ again. Based on the mode status matrix, we obtain $\left\vert
1\right\rangle \otimes \left\vert 0\right\rangle \otimes \left\vert
0\right\rangle $, $\left\vert 0\right\rangle \otimes \left\vert
0\right\rangle \otimes \left\vert 1\right\rangle $, $\left\vert
0\right\rangle \otimes \left\vert 1\right\rangle \otimes \left\vert
0\right\rangle $ for the selection of $R_{1}$, $R_{2}$, $R_{3}$,
respectively. We find an interesting fact that when we obtain the state $%
\left\vert 1\right\rangle $ of the first field, $R_{1}$ must be selected,
thus only the $\left\vert 0\right\rangle \otimes \left\vert 0\right\rangle $
state can be obtained from the other two fields; otherwise when we obtain
the state $\left\vert 0\right\rangle $ of the first field, the selection can
be $R_{2}$ or $R_{3}$, thus the state of $\left\vert 0\right\rangle \otimes
\left\vert 1\right\rangle +\left\vert 1\right\rangle \otimes \left\vert
0\right\rangle $ can be obtained from the other two fields. This fact is
quite similar with the case of quantum measurement and the collapse
phenomenon for W state in quantum mechanics. We can simulate the state based
on the mode status matrix expect a normalization factor, 
\begin{eqnarray}
\left\vert \Psi _{W}\right\rangle &=&\left( \tilde{\alpha}_{1}^{1}\left\vert
0\right\rangle +\tilde{\beta}_{1}^{1}\left\vert 1\right\rangle \right)
\otimes \left( \tilde{\alpha}_{2}^{2}\left\vert 0\right\rangle +\tilde{\beta}%
_{2}^{2}\left\vert 1\right\rangle \right) \otimes \left( \tilde{\alpha}%
_{3}^{3}\left\vert 0\right\rangle +\tilde{\beta}_{3}^{3}\left\vert
1\right\rangle \right)  \label{eq38} \\
&&+\left( \tilde{\alpha}_{1}^{2}\left\vert 0\right\rangle +\tilde{\beta}%
_{1}^{2}\left\vert 1\right\rangle \right) \otimes \left( \tilde{\alpha}%
_{2}^{3}\left\vert 0\right\rangle +\tilde{\beta}_{2}^{3}\left\vert
1\right\rangle \right) \otimes \left( \tilde{\alpha}_{3}^{1}\left\vert
0\right\rangle +\tilde{\beta}_{3}^{1}\left\vert 1\right\rangle \right) 
\nonumber \\
&&+\left( \tilde{\alpha}_{1}^{3}\left\vert 0\right\rangle +\tilde{\beta}%
_{1}^{3}\left\vert 1\right\rangle \right) \otimes \left( \tilde{\alpha}%
_{2}^{1}\left\vert 0\right\rangle +\tilde{\beta}_{2}^{1}\left\vert
1\right\rangle \right) \otimes \left( \tilde{\alpha}_{3}^{2}\left\vert
0\right\rangle +\tilde{\beta}_{3}^{2}\left\vert 1\right\rangle \right) 
\nonumber \\
&=&\left\vert 100\right\rangle +\left\vert 010\right\rangle +\left\vert
001\right\rangle .  \nonumber
\end{eqnarray}

Above we discussed the possibility of simulating several typical quantum
states by classical fields and mechanism of coherent detection.

\section{Simulation of quantum computation \label{sec4}}

In this chapter we will propose a method to simulate quantum computation. In
quantum computation, any quantum state can be obtained from initial states
by using unitary transformation of universal gates. Similarly, we can
construct simulation of all kinds of quantum states by using a gate array,
such as GHZ state and W state, even to realize Shor's algorithm and Grover's
algorithm.

\subsection{Gate array model to simulate quantum computation \label{sec4.1}}

\begin{figure}[htbp]
\centering\includegraphics[width=4.804in]{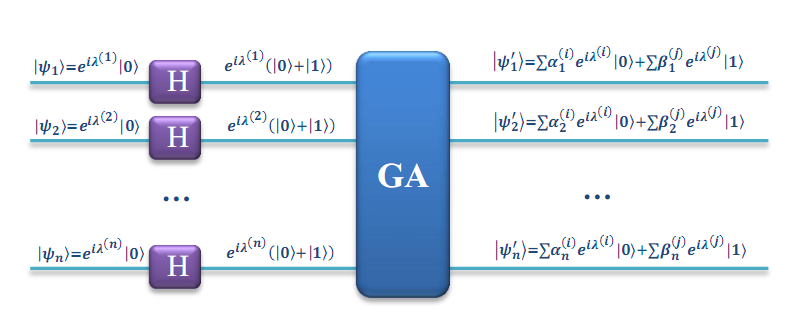}
\caption{A gate array model to simulate quantum computation is shown, where
GA denotes the gate array.}
\label{7}
\end{figure}

In Ref. \cite{Fu1}, a constructure pathway of simulation states is shown.
Here we use the same model as shown Fig. \ref{7}, however a gate array (GA)
is proposed instead of the unitary transformation to produce simulation
states. Now we discuss some basic units of gate array model.

(1) Combiner and splitter

\begin{figure}[htbp]
\centering\includegraphics[width=2.1421in]{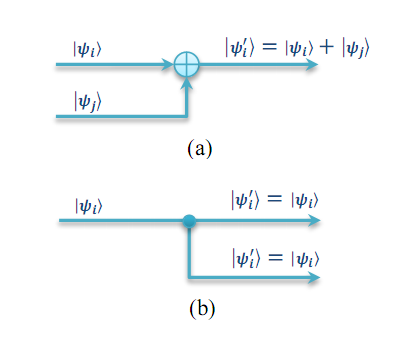}
\caption{Two basic devices (a) combiner and (b) splitter are shown.}
\label{8}
\end{figure}

Different from quantum state, we can conveniently combine and split a
classical field by using an optical splitter device. Therefore we define two
basic device as combiner and splitter as shown in Fig. \ref{8} (a) and (b),
respectively.

(b) Mode control Gates

\begin{figure}[htbp]
\centering\includegraphics[width=3.314in]{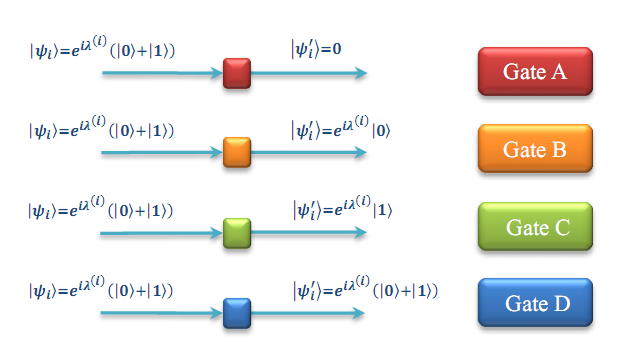}
\caption{Mode control gates as selective mode transit devices are shown.}
\label{9}
\end{figure}

Further, we define $4$ kinds of mode control gates as selective mode transit
devices with one input and one output as shown in Fig. \ref{9}. They are
defined by following: 
\begin{eqnarray}
\mathbf{Gate}A &:&\left\vert \psi _{i}\right\rangle =e^{i\lambda ^{\left(
i\right) }}\left( \left\vert 0\right\rangle +\left\vert 1\right\rangle
\right) \rightarrow \left\vert \psi _{i}^{\prime }\right\rangle =0,
\label{eq39} \\
\mathbf{Gate}B &:&\left\vert \psi _{i}\right\rangle =e^{i\lambda ^{\left(
i\right) }}\left( \left\vert 0\right\rangle +\left\vert 1\right\rangle
\right) \rightarrow \left\vert \psi _{i}^{\prime }\right\rangle =e^{i\lambda
^{\left( i\right) }}\left\vert 0\right\rangle ,  \nonumber \\
\mathbf{Gate}C &:&\left\vert \psi _{i}\right\rangle =e^{i\lambda ^{\left(
i\right) }}\left( \left\vert 0\right\rangle +\left\vert 1\right\rangle
\right) \rightarrow \left\vert \psi _{i}^{\prime }\right\rangle =e^{i\lambda
^{\left( i\right) }}\left\vert 1\right\rangle ,  \nonumber \\
\mathbf{Gate}D &:&\left\vert \psi _{i}\right\rangle =e^{i\lambda ^{\left(
i\right) }}\left( \left\vert 0\right\rangle +\left\vert 1\right\rangle
\right) \rightarrow \left\vert \psi _{i}^{\prime }\right\rangle =e^{i\lambda
^{\left( i\right) }}\left( \left\vert 0\right\rangle +\left\vert
1\right\rangle \right) .  \nonumber
\end{eqnarray}

\begin{figure}[htbp]
\centering\includegraphics[width=4.3656in]{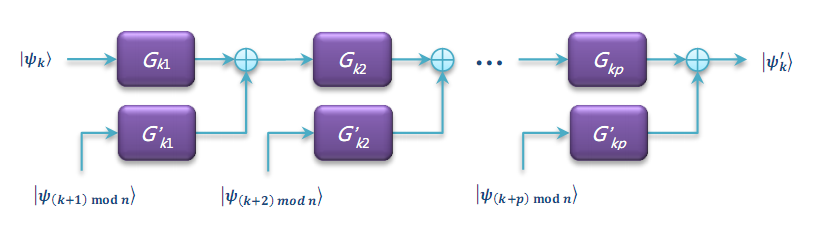}
\caption{A basic structrue of gate array model is shown.}
\label{10}
\end{figure}

Now we will discuss a basic structure of gate array model. According to Sec. %
\ref{sec3.3.2}, we can simulation of quantum states by using sequence
permutations. Similar to field programmable gate array (FPGA), we propose a
simple structure of gate array to obtain permutation structure as shown in
Fig. \ref{10}. Gates $G_{kj}$ constituted by the basic units can transform $%
\left\vert \psi _{k}\right\rangle $ to achieve certain $\left\vert \psi
_{k}^{\prime }\right\rangle $. It is easy to know that a sequence
permutation with circulation of $p+1$ needs at least $p$ combiner devices
and $2p$ control gates. Any states is capable to be constructed by applying
this structure that will be strictly proved in future paper.

\begin{figure}[htbp]
\centering\includegraphics[width=6.3088in]{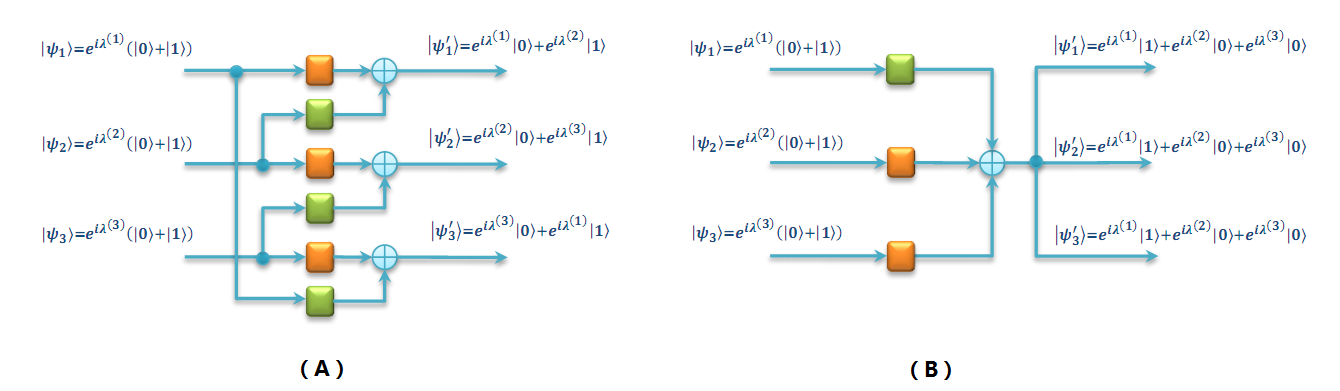}
\caption{Gate array models to transform product states to (a) GHZ state and
(b) W state are shown.}
\label{11}
\end{figure}

Finally, we illustrate two gate array models to transform product states to
GHZ state and W state as shown in Fig. \ref{11}.

\subsection{Simulation of quantum algorithm \label{sec4.2}}

\subsubsection{Shor's Algorithm \label{sec4.2.1}}

Shor's algorithm is a quantum algorithm for integer factorization that runs
only in polynomial time on a quantum computer \cite{Shor,Shor2}.
Specifically it takes time and quantum gates of order $O((\log N)^{2}(\log
\log N)(\log \log \log N))$ using fast multiplication, demonstrating that
the integer factorization problem can be efficiently solved on a quantum
computer and is thus in the complexity class BQP.

In this chapter, we discuss the simulation of Shor's algorithm. A algorithm
similar Shor's algorithm is demonstrated, that factored $N=15$ into $3\times
5$, using the gate array model with $8$ classical optical fields. First, we
chose a random number $a$ coprime with $15$, for example $a=7$. We define a
function as followed 
\begin{equation}
f\left( x\right) =a^{x} mod N=7^{x} mod 15.  \label{eq40}
\end{equation}%
The key step of the Shor's algorithm is to obtain the period $r$ to satisfy 
\begin{equation}
f\left( x+r\right) =7^{x+r} mod 15=7^{x} mod 15=f\left( x\right) .
\label{eq41}
\end{equation}%
In order to construct $f\left( x\right) $, we prepare $8$ classical optical
fields modulated with $8$ PPSs: 
\begin{equation}
\left\vert \psi _{k}\right\rangle =e^{i\lambda ^{\left( k\right) }}\left(
\left\vert 0\right\rangle +\left\vert 1\right\rangle \right) ,k=1\ldots 8.
\label{eq42}
\end{equation}%
We can express the product state as followed: 
\begin{equation}
\left\vert \Psi \right\rangle =\left\vert \psi _{1}\right\rangle \otimes
\ldots \otimes \left\vert \psi _{8}\right\rangle
=\sum\limits_{j=0}^{255}\left\vert j\right\rangle .  \label{eq43}
\end{equation}

\begin{figure}[htbp]
\centering\includegraphics[width=5.0704in]{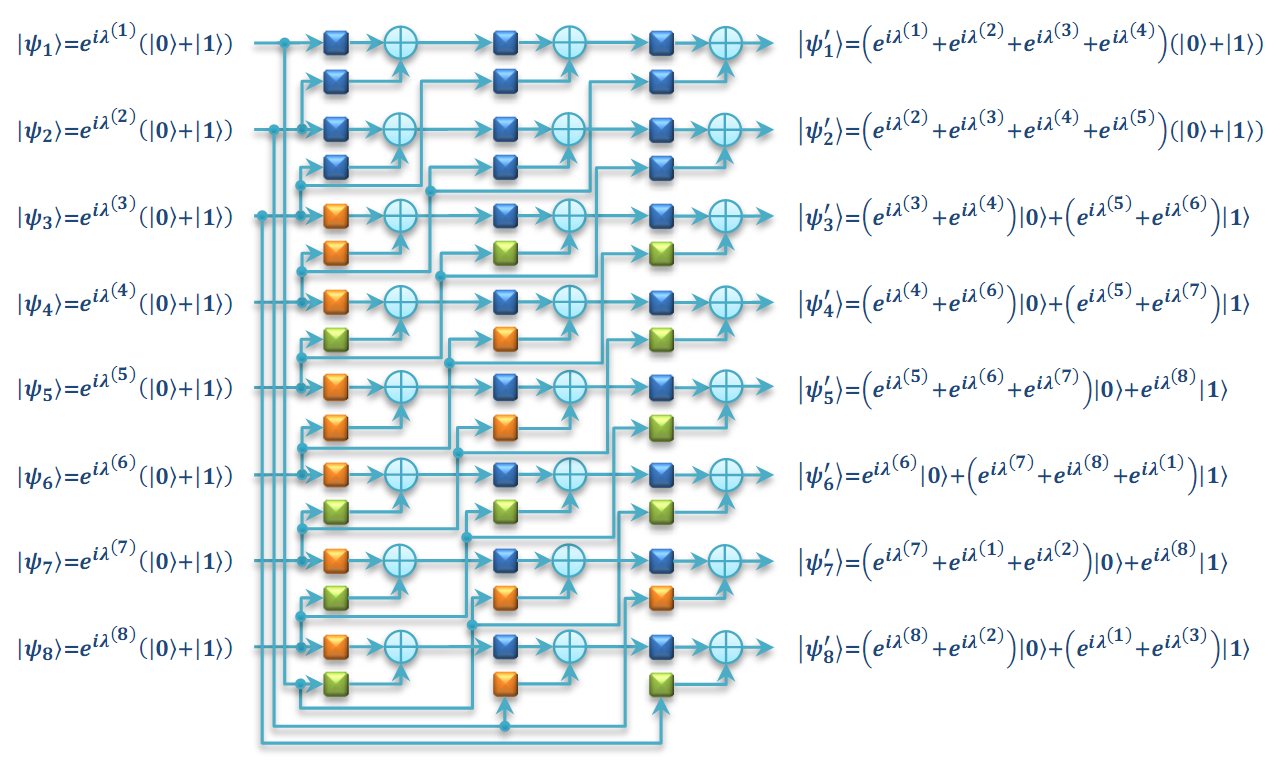}
\caption{A gate array model to realize Shor's algorithm is shown.}
\label{12}
\end{figure}

Further, we construct a gate array model as shown in Fig. \ref{12}. After
passing through the gate array and the classical fields will become the
forms as follows: 
\begin{eqnarray}
\left\vert \psi _{1}^{\prime }\right\rangle &=&\left( e^{i\lambda ^{\left(
1\right) }}+e^{i\lambda ^{\left( 2\right) }}+e^{i\lambda ^{\left( 3\right)
}}+e^{i\lambda ^{\left( 4\right) }}\right) \left( \left\vert 0\right\rangle
+\left\vert 1\right\rangle \right) ,  \label{eq44} \\
\left\vert \psi _{2}^{\prime }\right\rangle &=&\left( e^{i\lambda ^{\left(
2\right) }}+e^{i\lambda ^{\left( 3\right) }}+e^{i\lambda ^{\left( 4\right)
}}+e^{i\lambda ^{\left( 5\right) }}\right) \left( \left\vert 0\right\rangle
+\left\vert 1\right\rangle \right) ,  \nonumber \\
\left\vert \psi _{3}^{\prime }\right\rangle &=&\left( e^{i\lambda ^{\left(
3\right) }}+e^{i\lambda ^{\left( 4\right) }}\right) \left\vert
0\right\rangle +\left( e^{i\lambda ^{\left( 5\right) }}+e^{i\lambda ^{\left(
6\right) }}\right) \left\vert 1\right\rangle ,  \nonumber \\
\left\vert \psi _{4}^{\prime }\right\rangle &=&\left( e^{i\lambda ^{\left(
4\right) }}+e^{i\lambda ^{\left( 6\right) }}\right) \left\vert
0\right\rangle +\left( e^{i\lambda ^{\left( 5\right) }}+e^{i\lambda ^{\left(
7\right) }}\right) \left\vert 1\right\rangle ,  \nonumber \\
\left\vert \psi _{5}^{\prime }\right\rangle &=&\left( e^{i\lambda ^{\left(
5\right) }}+e^{i\lambda ^{\left( 6\right) }}+e^{i\lambda ^{\left( 7\right)
}}\right) \left\vert 0\right\rangle +e^{i\lambda ^{\left( 8\right)
}}\left\vert 1\right\rangle ,  \nonumber \\
\left\vert \psi _{6}^{\prime }\right\rangle &=&e^{i\lambda ^{\left( 6\right)
}}\left\vert 0\right\rangle +\left( e^{i\lambda ^{\left( 7\right)
}}+e^{i\lambda ^{\left( 8\right) }}+e^{i\lambda ^{\left( 1\right) }}\right)
\left\vert 1\right\rangle ,  \nonumber \\
\left\vert \psi _{7}^{\prime }\right\rangle &=&\left( e^{i\lambda ^{\left(
7\right) }}+e^{i\lambda ^{\left( 1\right) }}+e^{i\lambda ^{\left( 2\right)
}}\right) \left\vert 0\right\rangle +e^{i\lambda ^{\left( 8\right)
}}\left\vert 1\right\rangle ,  \nonumber \\
\left\vert \psi _{8}^{\prime }\right\rangle &=&\left( e^{i\lambda ^{\left(
8\right) }}+e^{i\lambda ^{\left( 2\right) }}\right) \left\vert
0\right\rangle +\left( e^{i\lambda ^{\left( 1\right) }}+e^{i\lambda ^{\left(
3\right) }}\right) \left\vert 1\right\rangle .  \nonumber
\end{eqnarray}

By using the quadrature demodulation, we can obtain the mode status matrix: 
\begin{equation}
M\left( \tilde{\alpha}_{i}^{j},\tilde{\beta}_{i}^{j}\right) =\left( 
\begin{array}{cccccccc}
\left( 1,1\right) & \left( 1,1\right) & \left( 1,1\right) & \left( 1,1\right)
& 0 & 0 & 0 & 0 \\ 
0 & \left( 1,1\right) & \left( 1,1\right) & \left( 1,1\right) & \left(
1,1\right) & 0 & 0 & 0 \\ 
0 & 0 & \left( 1,0\right) & \left( 1,0\right) & \left( 0,1\right) & \left(
0,1\right) & 0 & 0 \\ 
0 & 0 & 0 & \left( 1,0\right) & \left( 0,1\right) & \left( 1,0\right) & 
\left( 0,1\right) & 0 \\ 
0 & 0 & 0 & 0 & \left( 1,0\right) & \left( 1,0\right) & \left( 1,0\right) & 
\left( 0,1\right) \\ 
\left( 0,1\right) & 0 & 0 & 0 & 0 & \left( 1,0\right) & \left( 0,1\right) & 
\left( 0,1\right) \\ 
\left( 1,0\right) & \left( 1,0\right) & 0 & 0 & 0 & 0 & \left( 1,0\right) & 
\left( 0,1\right) \\ 
\left( 0,1\right) & \left( 1,0\right) & \left( 0,1\right) & 0 & 0 & 0 & 0 & 
\left( 1,0\right)%
\end{array}%
\right)  \label{eq45}
\end{equation}%
Using the scheme mentioned in Sec. \ref{sec3.2}, we can obtain the simulated
states: 
\begin{eqnarray}
\left\vert \Psi ^{\prime }\right\rangle &=&\left( \left\vert 0\right\rangle
+\left\vert 4\right\rangle +\left\vert 8\right\rangle +\left\vert
12\right\rangle \right) \left\vert 1\right\rangle  \label{eq46} \\
&&+\left( \left\vert 1\right\rangle +\left\vert 5\right\rangle +\left\vert
9\right\rangle +\left\vert 13\right\rangle \right) \left\vert 7\right\rangle
\nonumber \\
&&+\left( \left\vert 2\right\rangle +\left\vert 6\right\rangle +\left\vert
10\right\rangle +\left\vert 14\right\rangle \right) \left\vert 4\right\rangle
\nonumber \\
&&+\left( \left\vert 3\right\rangle +\left\vert 7\right\rangle +\left\vert
11\right\rangle +\left\vert 15\right\rangle \right) \left\vert
13\right\rangle .  \nonumber
\end{eqnarray}

There are four kinds of superposition classified from last four qubits
containing the values of $f\left( x\right) $ ($\left\vert 1\right\rangle
,\left\vert 7\right\rangle ,\left\vert 4\right\rangle $ and $\left\vert
13\right\rangle $) in output states, which means the period of $f\left(
x\right) =7^{x} mod 15$ is $r=4$. It is worth noting that, different from
quantum computing, we obtain the expected period of without operating
quantum Fourier transformation. The remaining task is much easier. Because $%
N=15,a=7,r=4$, we obtain 
\begin{equation}
\gcd \left( a^{\frac{r}{2}}\pm 1,N\right) =\gcd \left( 7^{\frac{r}{2}}\pm
1,15\right) =\gcd \left( 49\pm 1,15\right) ,  \label{eq47}
\end{equation}%
where $\gcd (48,15)=3$, and $\gcd (50,15)=5$. Finally, we can deduced that $%
N\left( 15\right) =3\times 5$.

Here, this is the simulation of quantum Shor's algorithm. Apparently, this
algorithm obtains the period of $f\left( x\right) $ without quantum fourier
transform and takes time and gates of order $O(\left( \log N\right) ^{2})$
by rough estimation. We will discuss this problem in detail in future paper.

\subsubsection{Grover's Algorithm \label{sec4.2.2}}

Grover's algorithm is a quantum algorithm for searching an unsorted database
with $N$ entries in $O(N^{1/2})$ time and using $O(\log N)$ storage space 
\cite{Grover}. Grover's algorithm demostrates that in the quantum model
searching can be done faster than classical computation; in fact its time
complexity $O(N^{1/2})$ is asymptotically the fastest possible for searching
an unsorted database in the linear quantum model. However, it only provides
a quadratic speedup rather than exponential speedup over their classical
counterparts.

We now discuss classical simulation of Grover's algorithm. According to
Grover's algorithm, the steps of the simulation are given as follows. Let $%
\left\vert S\right\rangle $ denote the uniform superposition over $N$
states, 
\begin{equation}
\left\vert S\right\rangle =\sum\limits_{i=1}^{N}\left\vert x_{i}\right\rangle
\label{eq48}
\end{equation}%
where $\left\vert x_{i}\right\rangle $ are the states. For example, we
obtain $\left\vert S\right\rangle $ as a superposition state of $13$ random
numbers: 
\begin{eqnarray}
\left\vert S\right\rangle &=&\left\vert 61\right\rangle +\left\vert
63\right\rangle +\left\vert 117\right\rangle +\left\vert 125\right\rangle
+\left\vert 140\right\rangle +\left\vert 142\right\rangle +\left\vert
148\right\rangle  \label{eq49} \\
&&+\left\vert 212\right\rangle +\left\vert 187\right\rangle +\left\vert
59\right\rangle +\left\vert 238\right\rangle +\left\vert 247\right\rangle
+\left\vert 76\right\rangle  \nonumber \\
&=&\left\vert 00111101\right\rangle +\left\vert 00111111\right\rangle
+\left\vert 01110101\right\rangle +\left\vert 01111101\right\rangle 
\nonumber \\
&&+\left\vert 10001100\right\rangle +\left\vert 10001110\right\rangle
+\left\vert 10010100\right\rangle +\left\vert 11010100\right\rangle 
\nonumber \\
&&+\left\vert 10111011\right\rangle +\left\vert 00111011\right\rangle
+\left\vert 11101110\right\rangle +\left\vert 11110111\right\rangle 
\nonumber \\
&&+\left\vert 01001100\right\rangle .  \nonumber
\end{eqnarray}%
We choose $8$ classical fields modulated with $8$ PPSs and after passing
through a suitable gate array, that become the forms as follows: 
\begin{eqnarray}
\left\vert \psi _{1}^{\prime }\right\rangle &=&\left( e^{i\lambda ^{\left(
1\right) }}+e^{i\lambda ^{\left( 2\right) }}+e^{i\lambda ^{\left( 5\right)
}}+e^{i\lambda ^{\left( 8\right) }}\right) \left\vert 0\right\rangle +\left(
e^{i\lambda ^{\left( 3\right) }}+e^{i\lambda ^{\left( 4\right)
}}+e^{i\lambda ^{\left( 5\right) }}+e^{i\lambda ^{\left( 6\right)
}}+e^{i\lambda ^{\left( 7\right) }}\right) \left\vert 1\right\rangle , \\
\left\vert \psi _{2}^{\prime }\right\rangle &=&\left( e^{i\lambda ^{\left(
2\right) }}+e^{i\lambda ^{\left( 4\right) }}+e^{i\lambda ^{\left( 5\right)
}}+e^{i\lambda ^{\left( 6\right) }}\right) \left\vert 0\right\rangle +\left(
e^{i\lambda ^{\left( 1\right) }}+e^{i\lambda ^{\left( 3\right)
}}+e^{i\lambda ^{\left( 5\right) }}+e^{i\lambda ^{\left( 7\right)
}}+e^{i\lambda ^{\left( 8\right) }}\right) \left\vert 1\right\rangle , 
\nonumber \\
\left\vert \psi _{3}^{\prime }\right\rangle &=&\left( e^{i\lambda ^{\left(
2\right) }}+e^{i\lambda ^{\left( 5\right) }}+e^{i\lambda ^{\left( 6\right)
}}\right) \left\vert 0\right\rangle +\left( e^{i\lambda ^{\left( 1\right)
}}+e^{i\lambda ^{\left( 3\right) }}+e^{i\lambda ^{\left( 4\right)
}}+e^{i\lambda ^{\left( 7\right) }}+e^{i\lambda ^{\left( 8\right) }}\right)
\left\vert 1\right\rangle ,  \nonumber \\
\left\vert \psi _{4}^{\prime }\right\rangle &=&\left( e^{i\lambda ^{\left(
1\right) }}+e^{i\lambda ^{\left( 3\right) }}+e^{i\lambda ^{\left( 6\right)
}}+e^{i\lambda ^{\left( 8\right) }}\right) \left\vert 0\right\rangle +\left(
e^{i\lambda ^{\left( 2\right) }}+e^{i\lambda ^{\left( 4\right)
}}+e^{i\lambda ^{\left( 5\right) }}+e^{i\lambda ^{\left( 7\right) }}\right)
\left\vert 1\right\rangle ,  \nonumber \\
\left\vert \psi _{5}^{\prime }\right\rangle &=&\left( e^{i\lambda ^{\left(
3\right) }}+e^{i\lambda ^{\left( 6\right) }}+e^{i\lambda ^{\left( 8\right)
}}\right) \left\vert 0\right\rangle +\left( e^{i\lambda ^{\left( 1\right)
}}+e^{i\lambda ^{\left( 2\right) }}+e^{i\lambda ^{\left( 4\right)
}}+e^{i\lambda ^{\left( 5\right) }}+e^{i\lambda ^{\left( 6\right)
}}+e^{i\lambda ^{\left( 7\right) }}\right) \left\vert 1\right\rangle , 
\nonumber \\
\left\vert \psi _{6}^{\prime }\right\rangle &=&e^{i\lambda ^{\left( 2\right)
}}\left\vert 0\right\rangle +\left( e^{i\lambda ^{\left( 1\right)
}}+e^{i\lambda ^{\left( 3\right) }}+e^{i\lambda ^{\left( 4\right)
}}+e^{i\lambda ^{\left( 5\right) }}+e^{i\lambda ^{\left( 6\right)
}}+e^{i\lambda ^{\left( 7\right) }}+e^{i\lambda ^{\left( 8\right) }}\right)
\left\vert 1\right\rangle ,  \nonumber \\
\left\vert \psi _{7}^{\prime }\right\rangle &=&\left( e^{i\lambda ^{\left(
1\right) }}+e^{i\lambda ^{\left( 2\right) }}+e^{i\lambda ^{\left( 6\right)
}}+e^{i\lambda ^{\left( 7\right) }}+e^{i\lambda ^{\left( 8\right) }}\right)
\left\vert 0\right\rangle +\left( e^{i\lambda ^{\left( 1\right)
}}+e^{i\lambda ^{\left( 3\right) }}+e^{i\lambda ^{\left( 4\right)
}}+e^{i\lambda ^{\left( 5\right) }}+e^{i\lambda ^{\left( 7\right) }}\right)
\left\vert 1\right\rangle ,  \nonumber \\
\left\vert \psi _{8}^{\prime }\right\rangle &=&\left( e^{i\lambda ^{\left(
2\right) }}+e^{i\lambda ^{\left( 3\right) }}+e^{i\lambda ^{\left( 5\right)
}}+e^{i\lambda ^{\left( 7\right) }}\right) \left\vert 0\right\rangle +\left(
e^{i\lambda ^{\left( 1\right) }}+e^{i\lambda ^{\left( 4\right)
}}+e^{i\lambda ^{\left( 6\right) }}+e^{i\lambda ^{\left( 8\right) }}\right)
\left\vert 1\right\rangle .  \nonumber
\end{eqnarray}

Due to the scheme mentioned in Sec. \ref{sec3.2}, any simulated state must
correspond to a certain sequence permutation. Therefore, the problem to
determine whether $\left\vert x\right\rangle $ exists in $\left\vert
S\right\rangle $ become that to search the corresponding sequence
permutation. For example, we search the number $\left\vert x\right\rangle
=\left\vert 148\right\rangle =\left\vert 10010100\right\rangle $ in $%
\left\vert S\right\rangle $. First, the classical fields of state $%
\left\vert S\right\rangle $ pass through a gate array as shown in Fig. \ref%
{13}. Then we obtain the mode status matrix by using the quadrature
demodulation: 
\begin{equation}
M\left( \tilde{\alpha}_{i}^{j},\tilde{\beta}_{i}^{j}\right) =\left( 
\begin{array}{cccccccc}
0 & 0 & \left( 0,1\right) & \left( 0,1\right) & \left( 0,1\right) & \left(
0,1\right) & \left( 0,1\right) & 0 \\ 
0 & \left( 1,0\right) & 0 & \left( 1,0\right) & \left( 1,0\right) & \left(
1,0\right) & 0 & 0 \\ 
0 & \left( 1,0\right) & 0 & 0 & \left( 1,0\right) & \left( 1,0\right) & 0 & 0
\\ 
0 & \left( 0,1\right) & 0 & \left( 0,1\right) & \left( 0,1\right) & 0 & 
\left( 0,1\right) & 0 \\ 
0 & 0 & \left( 1,0\right) & 0 & 0 & \left( 1,0\right) & 0 & \left( 1,0\right)
\\ 
\left( 0,1\right) & 0 & \left( 0,1\right) & \left( 0,1\right) & \left(
0,1\right) & \left( 0,1\right) & \left( 0,1\right) & \left( 0,1\right) \\ 
\left( 1,0\right) & \left( 1,0\right) & 0 & 0 & 0 & \left( 1,0\right) & 
\left( 1,0\right) & \left( 1,0\right) \\ 
0 & \left( 1,0\right) & \left( 1,0\right) & 0 & \left( 1,0\right) & 0 & 
\left( 1,0\right) & 0%
\end{array}%
\right)
\end{equation}%
Finally, it is easy to search only corresponding sequence permutation $%
R_{4}=\left\{ \lambda ^{\left( 4\right) },\lambda ^{\left( 5\right)
},\lambda ^{\left( 6\right) },\lambda ^{\left( 7\right) },\lambda ^{\left(
8\right) },\lambda ^{\left( 1\right) },\lambda ^{\left( 2\right) },\lambda
^{\left( 3\right) }\right\} $. If we choose $\left\vert x\right\rangle
=\left\vert 240\right\rangle =\left\vert 11110000\right\rangle $, we can
obtain the mode status matrix 
\begin{equation}
M\left( \tilde{\alpha}_{i}^{j},\tilde{\beta}_{i}^{j}\right) =\left( 
\begin{array}{cccccccc}
0 & 0 & \left( 0,1\right) & \left( 0,1\right) & \left( 0,1\right) & \left(
0,1\right) & \left( 0,1\right) & 0 \\ 
\left( 0,1\right) & 0 & \left( 0,1\right) & 0 & \left( 0,1\right) & 0 & 
\left( 0,1\right) & \left( 0,1\right) \\ 
\left( 0,1\right) & 0 & \left( 0,1\right) & \left( 0,1\right) & 0 & 0 & 
\left( 0,1\right) & \left( 0,1\right) \\ 
0 & \left( 0,1\right) & 0 & \left( 0,1\right) & \left( 0,1\right) & 0 & 
\left( 0,1\right) & 0 \\ 
0 & 0 & \left( 1,0\right) & 0 & 0 & \left( 1,0\right) & 0 & \left( 1,0\right)
\\ 
0 & \left( 1,0\right) & 0 & 0 & 0 & 0 & 0 & 0 \\ 
\left( 1,0\right) & \left( 1,0\right) & 0 & 0 & 0 & \left( 1,0\right) & 
\left( 1,0\right) & \left( 1,0\right) \\ 
0 & \left( 1,0\right) & \left( 1,0\right) & 0 & \left( 1,0\right) & 0 & 
\left( 1,0\right) & 0%
\end{array}%
\right)
\end{equation}%
In the mode status matrix, we can not search any corresponding sequence
permutation. Therefore we can conclude $\left\vert x\right\rangle
=\left\vert 240\right\rangle $ does not exist in $\left\vert S\right\rangle $%
.

Here, this is the simulation of quantum Grover's algorithm. Different from
Grover's algorithm, this algorithm for searching an unsorted database with $%
N $ entries in $O(\left( \log N\right) ^{2})$ time and using $O(\left( \log
N\right) ^{2})$ storage space by rough estimation. We will discuss this
problem in detail also in future paper.

\begin{figure}[htbp]
\centering\includegraphics[width=4.7703in]{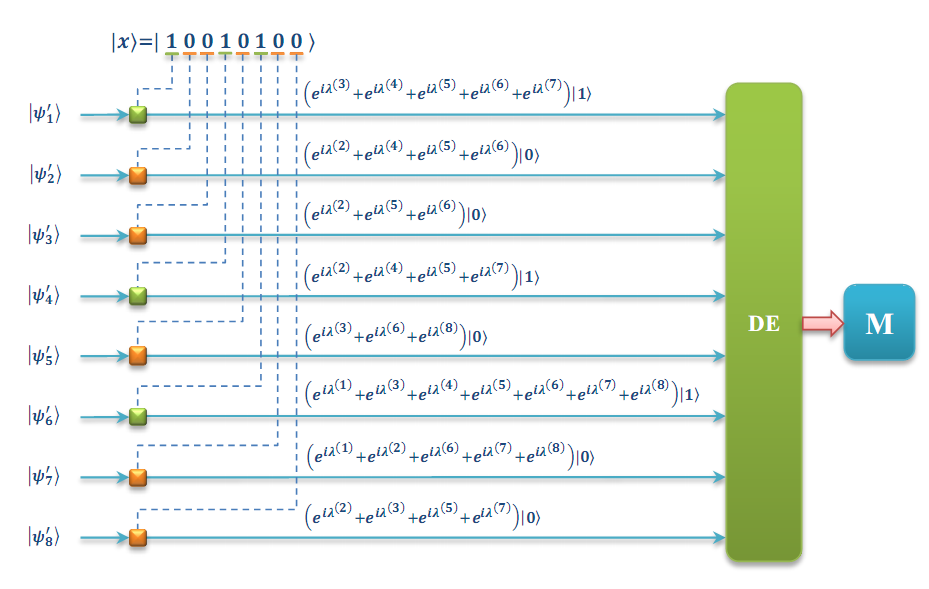}
\caption{A gate array model to select $\left\vert x\right\rangle $ from $%
\left\vert S\right\rangle $ is shown.}
\label{13}
\end{figure}

\section{Conclusions \label{sec5}}

In this paper, we have discussed a new scheme to simulate quantum states by
using classical optical fields modulated with pseudorandom phase sequences.
We first demonstrated that $n$ classical optical fields modulated with $n$\
different PPSs can constitute a $n2^{n}$-dimensional Hilbert space that
contains tensor product structure similar to quantum systems. Further, by
performing quadrature demodulation scheme, we obtained the mode status
matrix of the simulating classical optical fields, based on which we
proposed a sequence permutation mechanism to simulate the simulated quantum
states\textbf{.} Besides, classical simulation of several typical quantum
states was discussed, including product state, Bell states, GHZ state and W
state. Finally, we generalized our simulation and discussed a generalized
gate array model to simulate quantum computation. The research on simulation
of quantum states is important, for it not only provides useful insights
into fundamental features of quantum mechanics, but also yields new insights
into quantum computation and quantum communication.

\begin{description}
\item \newpage

\item[Fig. 1] The time sequence relationship of the PPS is shown.

\item[Fig. 2] The PPS encoding scheme for one input field is shown, where
PNG denotes the pseudorandom number generator and PM denotes the phase
modulator.

\item[Fig. 3] The PPS quadrature demodulation scheme for one input field
with single mode $e^{i\lambda ^{\left( i\right) }}\left\vert 0\right\rangle $
or $e^{i\lambda ^{\left( i\right) }}\left\vert 1\right\rangle $ is shown.

\item[Fig. 4] The PPS quadrature demodulation scheme for one field with two
orthogonal modes is shown, where the blue block MS denote the mode splitter.

\item[Fig. 5] The PPS quadrature demodulation scheme for multiple input
fields is shown, where the DE block is shown in Fig. \ref{4}.

\item[Fig. 6] The scheme to simulate quantum state is shown, the mode status
matrix $M\left( \tilde{\alpha}_{i}^{j},\tilde{\beta}_{i}^{j}\right) $
related to $i$th classical optical field and the reference PPS $\lambda
^{\left( j\right) }$, in which the mode status with same color for same
sequence permutation.

\item[Fig. 7] A gate array model to simulate quantum computation is shown,
where GA denotes the gate array.

\item[Fig. 8] Two basic devices (a) combiner and (b) splitter are shown.

\item[Fig. 9] Mode control gates as selective mode transit devices are shown.

\item[Fig. 10] A basic structure of gate array model is shown.

\item[Fig. 11] Gate array models to transform product states to (a) GHZ
state and (b) W state are shown.

\item[Fig. 12] A gate array model to realize Shor's algorithm is shown.

\item[Fig. 13] A gate array model to select $\left\vert x\right\rangle $
from $\left\vert S\right\rangle $ is shown.
\end{description}

\end{document}